\documentclass[12pt,twoside]{article} 
\usepackage[english]{babel}
\usepackage{graphicx}
\usepackage{amsmath,amssymb,amsfonts}
\usepackage{mathtools,slashed,braket,tensor,accents}
\usepackage[makeroom]{cancel}
\usepackage{ifthen}
\usepackage{setspace}
\usepackage[utf8]{inputenc}
\usepackage{soul}
\usepackage[misc]{ifsym}
\usepackage[normalem]{ulem}
\usepackage{wrapfig}
\usepackage{lipsum} % For generating dummy text, can be removed in your actual document
\usepackage{float}
\usepackage{subfloat}
\usepackage[export]{adjustbox}
\usepackage{tikz}
\usepackage[textsize=tiny,color=red]{todonotes}

 \usepackage{kotex}%Korean
\usepackage{CJKutf8}
\usepackage[font=small, labelfont=bf]{caption}
\usepackage{subcaption, comment}
 \usepackage[]{titlesec} %you can choose e.g. [indentafter]
 \titleformat{name=\section}[block]{}{\thesection.}{0.8em}{\centering\bfseries}
 \titleformat{name=\subsection}[runin]{}{\thesubsection.}{0.5em}{\bfseries}[.]
\titleformat{name=\subsubsection}[runin]{}{\thetitle.}{0.5em}{\itshape}[.]

\usepackage[top=2.5cm, bottom=2.5cm, left=2.5cm, right=2.5cm]{geometry} 
\usepackage{setspace} 
\usepackage{color}
\definecolor{lighterblue}{RGB}{120,140,250}
\definecolor{lightblue}{RGB}{50,120,150}
\definecolor{darkred}{RGB}{190,0,0}
\definecolor{darkerred}{RGB}{170,0,0}
\definecolor{darkblue}{rgb}{0.0, 0.0, 0.5}
\definecolor{amber}{rgb}{1.0, 0.75,0.0}
\definecolor{arsenic}{RGB}{50,50,90}
\definecolor{goodgreen}{rgb}{0.0, 0.55, 0.2}
\definecolor{darkorange}{RGB}{255,100,0}
\definecolor{antiquefuchsia}{rgb}{0.57, 0.36, 0.51}
\definecolor{groen}{RGB}{50,156,105}
\definecolor{paarsblauw}{RGB}{92,86,209} %#5C56D1

\usepackage[colorlinks=true,
            linkcolor=groen,
            urlcolor=lightblue,
            citecolor=paarsblauw
            ]{hyperref}      

 \usepackage[
 maxcitenames=3,
 sorting=ynt,
 uniquename=false,
 backend=biber,
 uniquelist=false,
 sortcites=true,
 natbib=true,
 doi=false,
 url=false,
 isbn=false]{biblatex}
\DeclareNameAlias{sortname}{last-first}
\makeatletter
\renewcommand\hyper@natlinkbreak[2]{#1}
\makeatother
 \renewbibmacro{in:}{}
\DeclareNameAlias{sortname}{last-first}

 \DefineBibliographyStrings{english}{%
  byeditor = {Edited by},
}

 \newbibmacro{string+doiurlisbn}[1]{%
  \iffieldundef{doi}{%
    \iffieldundef{url}{%
      \iffieldundef{isbn}{%
        \iffieldundef{issn}{%
          #1%
        }{%
          \href{http://books.google.com/books?vid=ISSN\thefield{issn}}{#1}%
        }%
      }{%
        \href{http://books.google.com/books?vid=ISBN\thefield{isbn}}{#1}%
      }%
    }{%
      \href{\thefield{url}}{#1}%
    }%
  }{%
    \href{http://dx.doi.org/\thefield{doi}}{#1}%
  }%
}

\DeclareFieldFormat{title}{\usebibmacro{string+doiurlisbn}{\mkbibemph{#1}}}
\DeclareFieldFormat[article,incollection]{title}%
    {\usebibmacro{string+doiurlisbn}{\mkbibquote{#1}}}
  
\usepackage{tensind}
  \tensordelimiter{?}

\usepackage{thmtools} % Use 'restatable' to restate theorem in Appendix
\usepackage{thm-restate}

\newtheorem{prop}{Proposition}
\def\brack#1{ \left(   #1 \right) }

\def\blockbrack#1{ \left[   #1 \right] }

\newsavebox\MBox

\numberwithin{equation}{section}

\usepackage{xparse}
\usepackage{cleveref}

\usepackage{placeins}
\usepackage{float}

\date{December 10, 2025}
\begin{document}
% \title{\textsc{Conventionalism in general relativity: \\ 
% % Reichenbach’s 
% Weatherall $\&$ Manchak’s proof against theorem θ in context }}
\title{\textsc{Conventionalism in general relativity?: \\ 
 formal existence proofs and Reichenbach's theorem θ in context }}
 \author{Ruward Mulder\footnote{\textit{University of Cambridge, Trinity College, CB2 1TQ, Cambridge; and University of California Irvine, Logic and Philosophy of Science, 3151 Social Sciences Plaza A, Irvine CA 92697. \href{mailto:ramulder@uci.edu}{ramulder@uci.edu}. }}}
 \maketitle

\renewcommand{\baselinestretch}{1.} 

\vspace{-1cm}
{\footnotesize \begin{abstract}
\noindent {\scriptsize Weatherall and Manchak (2014) show that, under reasonable assumptions, Reichenbachean universal effects, constrained to a rank-2 tensor field representation in the geodesic equation, always exist in non-relativistic gravity but not so for relativistic spacetimes. Thus general relativity is less susceptible to underdetermination than its Newtonian predecessor. Dürr and Ben-Menahem (2022) argue these assumptions are exploitable as loopholes, effectively establishing a (rich) no-go theorem. I disambiguate between two targets of the proof, which have previously been conflated: the existence claim of at least one alternative geometry to a given one and Reichenbach's (in)famous ``theorem theta", which amounts to a universality claim that any geometry can function as an alternative to any other. I show there is no (rich) no-go theorem to save theorem theta. I illustrate this by explicitly breaking one of the assumptions and generalising the proof to torsionful spacetimes. Finally, I suggest a programmatic attitude: rather than undermining the proof one can use it to systematically and rigorously articulate stronger propositions to be proved, thereby systematically exploring the space of alternative spacetime theories. 
} 
\end{abstract}
}

% {\footnotesize \begin{abstract}
% \noindent {\scriptsize  Weatherall and Manchak (2014) show that Reichenbachean universal effects, constrained to a rank-2 tensor field representation in the geodesic equation, always exist in non-relativistic gravity but not so for relativistic spacetimes. Thus general relativity is less susceptible to underdetermination than its Newtonian predecessor. I reproduce this proof in more detail and pinpoint where assumptions are made. Dürr and Ben-Menahem (2022) argue these assumptions are exploitable as loopholes, effectively establishing a (rich) no-go theorem. I disambiguate between two targets of the proof, which have before been conflated: the existence claim of at least one alternative geometry to a given one and Reichenbach's (in)famous "theorem theta", which amounts to a universality claim that any geometry can function as an alternative to any other. I show there is no (rich) no-go theorem to save theorem theta. I illustrate this by explicitly breaking one of the assumptions and generalising the proof to torsionful spacetimes. Finally, I suggest a programmatic attitude: rather than undermining the proof one can use it to systematically and rigorously articulate stronger propositions to be proved, thereby systematically exploring the space of alternative spacetime theories. 
% } 
% \end{abstract}
% }

% \clearpage
% Table of Contents%
  \thispagestyle{empty}
 \renewcommand{\contentsname}{Table of contents}
  % \addcontentsline{toc}{section}{\protect\numberline{}Table of contents}
 % \footnotesize{
  %\tiny{
  \scriptsize{
{\hypersetup{linkcolor=black} \textsf{\tableofcontents}}
}
 % \thispagestyle{empty}
 
% \clearpage
\normalsize{ }
 \setstretch{1.0}
% \pagenumbering{arabic}

\thispagestyle{empty}
\clearpage 
 \setcounter{page}{1}

% \section{}

\section{Introduction} \label{sec:introConventionalism}
Conventionality about space, or spacetime, is the view in the epistemology of geometry
-- loosely associated with ideas by Poincaré, Schlick, Carnap, Reichenbach, and others --
that ascertaining the geometry of the world requires, in some way or another, a conventional aspect. 
Historically, this debate emerged against the backdrop of a nineteenth-century separation between mathematical and physical geometry: beginning with Riemann and culminating in Hilbert’s axiomatization, mathematical geometry became an abstract formal discipline, while physical geometry remained tied to empirically measurable magnitudes in an Helmholtzian sense. In this empiricist tradition, there is indeed an empirical question to be asked about the \textit{true} geometry of the world, as opposed to the purely formal (hence analytically true in Hilbert’s sense) structures of Euclidean and non-Euclidean geometry: physical practice is taken to reveal synthetic facts about the physical geometry of the world.\footnote{The literature is too vast to review, but a (non-exhaustive) list from which I draw is  
(\cite{%Grunbaum1963-GRNPPO, 
Sklar1974-SKLSTA, 
Dieks1987-DIEGAA, 
Dewar2022-DEWTEO-11,
Glymour,
Ben-Menahem2006-BENC-2, 
MilenaMatt,
Ivanova2015-IVACAW,
Ivanova2021-IVADAH,
% Wilholt2012-WILCPD,
DuerrRead2024,
EisenthalAbsolute, 
EisenthalPatton,
PesicGeometry,
Acuna2013-ACUAEO, 
% Dewar2020-DEWARW, 
Dewar2022Convention, 
Torretti,
FlaviaPadovaniReichi, 
Glymour1972Convention}). 
% Marschall2024}). 
} 

At the same time, geometric conventionalists hold that such an empiricist search for factual statements about physical geometry is beyond empirical reach: such statements can only be given empirical content by means of an anchoring point, a conventional choice that cannot itself be justified on empirical grounds. 
 Canonically, geometric conventionalism finds traction through the idea that different spacetime theories can be made empirically indistinguishable, despite being based on \textit{prima facie} distinct geometric structures. This can be achieved by introducing \textit{universal forces} (or \textit{universal effects}) that correct for the geometric differences of models with different geometries -- models within the same spacetime theory or between distinct spacetime theories -- in such a way that the empirically accessible trajectories of bodies remain unaffected. 
 % This possibility of a trade-off between forces and specific geometric structures 
% -- commonly understood in the tradition of physical geometry as having different physical correlates -- 
%  is then approached as a situation where we can choose freely between these models: 
% by convention. 
 % Thus, a realist interpretation of successful spacetime models providing us with true geometric properties of physical spacetime is threatened.   
%Hence one of the largest debates in the epistemology of spacetime since its conception: the conventionality of geometry.

Ostensibly, this is the position that J.O.~Weatherall and J.B.~Manchak (\cite*{WeatherallManchak2014}, henceforth: W$\&$M), set out to question rigorously in their \citetitle{WeatherallManchak2014}: 
% . They ask the central question: 
 \begin{quote}
If one understands ``force" in the standard way in the context of our best classical (i.e., non-quantum) theories of space and time, 
   can one accommodate different choices of geometry by postulating some sort of ``universal force field"? 
   (W$\&$M \cite*{WeatherallManchak2014}, p.~234). 
\end{quote}
They present two proofs, one non-relativistic and one for general relativity (GR),  restricting the universal force field to be represented by a rank-2 tensor field in the geodesic equation. %, set in the language of modern spacetime theories. 
 The non-relativistic proof \textit{confirms} that the conventionalist's claim that there are empirically adequate alternatives if one allows universal forces;  
the relativistic proof \textit{falsifies} the claim that this can always be done for general relativistic spacetimes. This paper will focus on the relativistic proof, which shows that 
   GR does not admit the same leeway for conventionalism as Newtonian gravity because there does not exist a reasonable `force' tensor that relates the geodesics of two conformally related metrics.  
 That is, \textit{under reasonable assumptions}. 

Certainly, W$\&$M showcase conventionalism in a novel way, mentioning much previous scholarship on the topic, but not obviously connecting to the questions posed in those works.   
An upshot is that they reformulate the problem in the full glory of modern spacetime theory, 
introducing much-needed rigour into a debate that has often relied on proofs of concept, thereby 
kicking off a fresh way to handle these questions. 
A downside is that at first sight their claims are remarkable, even shocking, given the numerous debates on this topic: has more than a century of scholarship missed a trick, 
perhaps because of the lack of a sufficiently rigorous analysis?
 In particular, a roadmap about whether we can now safely adopt an anti-conventionalist or realist conception of general relativistic spacetimes, or what kind of future work would be required to establish such commitment, 
is left to the reader. 

Given this unconventional anti-conventionality, 
controversy is to be expected. 
In fact, Patrick Dürr and Yemima Ben-Menahem (\cite*{DandbBM2022}, henceforth: D$\&$BM) write a response that makes no secret of their belief that the assumptions of the proof are
overly restrictive, 
not sufficiently justified, 
open to counterexamples, 
and not in accordance with historical and philosophical scholarship on the topic.
They present W$\&$M's relativistic proof as an inconsistency proof, lining up its assumptions together with an existence claim of alternative geometries 
 and pointing out how each of these assumptions can be rejected. As such, each rejection opens up possibilities for constructing empirically equivalent models.

 One can feel at a loss about the overall gain of this debate, if any.
 Have W$\&$M proven nothing of importance? 
 In this paper, I clarify two different kinds of underdetermination that serve to underpin two very different positions, each of a conventionalist but nonetheless distinct flavour.   
 Some conventionalists have in mind an existence claim: 
for any one given metric  
there exists \textit{at least one} other metric 
which can do the same empirically adequate job, given the universal forces (what I will call ($\text{\textbf{UDT}-}{\forall g \exists \tilde{g}}$) in \S\ref{sec:universalitynotexistence}). 
 Other conventionalists have in mind a universality claim ($\text{\textbf{UDT}-}{\forall g \forall \tilde{g}}$): \textit{any} metric can replace \textit{any} other, if suitably adjusted for. The most prominent universality claim was promulgated by Reichenbach (\cite*{ReichenbachSpaceTime}, p.~33, \textit{emphasis added}),    
  enshrined in his (in)famous \textit{theorem $\theta$}:
\begin{quote}
 \textit{Theorem $\theta$}: ``Given a geometry $G'$ to which the measuring instruments conform, we can imagine a universal force $F$ which affects the instruments in such a way that the actual geometry is an \textit{arbitrary} geometry $G$, while the observed deviation from $G$ is due to a universal deformation of the measuring instruments." 
\end{quote} 
\noindent This paper contextualises  W$\&$M's relativistic result by distinguishing between (restricted) existence and (restricted) universality claims.
Expanding on a list of assumptions of the proof identified by D$\&$BM, I discuss their loophole-based critique (\S\ref{sec:assumptionslisting}). 
Then, \S\ref{sec:universalitynotexistence} clarifies the structure of the proof by distinguishing between universality and existence claims. 
  I argue the proof (a) severely restricts but does not rule out underdetermination altogether, leaving an ever tightening space of potential alternative models, and (b) theorem $\theta$ is nevertheless successfully dispelled if universal effects are taken to behave as force fields in the standard way, without any (other) loopholes---an observation none of the authors addressed clearly. 
I then generalise the proof to torsionful spacetimes (\S\ref{sec:generlisetotorsion}), showing that (i) violating an assumption does not serve as a loophole to save theorem~$\theta$ in GR, and more generally that (ii) for a given torsionful connection there exists no conformally equivalent torsionful connection which can be so related. 
  Finally, \S\ref{sec:programmatic} proposes reframing W$\&$M’s assumptions as part of a research programme to develop stronger theorems, eventually culminating in a systematic grip on the formal landscape of the space of (conceived and unconceived) relativistic spacetime theories.

\section{Weatherall and Manchak's relativistic proof as a no-go theorem} \label{sec:assumptionslisting} 

 To mathematically cash out the question whether in GR one can accommodate different geometries via the postulation of a universal force field, W$\&$M pose it in terms of the affine-connection uniquely associated (given torsion-freeness and metric-compatibility) with the metric: 
for a relativistic spacetime on manifold $M$ with metric $g_{ab}$ with its associated Levi-Civita connection $\nabla$, and a distinct metric $\widetilde{g}_{ab}$ (on the same~$M$) with its associated connection $\widetilde{\nabla}$,  
and $\xi^a$ the velocity vector (the unit-norm, timelike vector tangent to a particle's curve $\gamma(\tau)$, i.e., $\xi^a=\dot{\gamma}$).
The question then reads: 
 ``Is there some rank-2 tensor field $?F_{ab}?$ such that, given a curve $\gamma$, $\gamma$ is a geodesic (up to reparametrisation) relative to $\nabla$ just in case its acceleration relative to $\widetilde{\nabla}$ is given by $?F^a_n?\tilde{\xi}^n$, where $\tilde{\xi}^a$ is the tangent field to $\gamma$ with unit length relative to $\tilde{g}_{ab}$?", providing  
  the ingredients sufficient to write the following proposition\footnote{%The names of the propositions as given by W$\&$M. 
 Proposition 1 is the non-relativistic case, which shows that a suitable force tensor \textit{does} exist. 
 % which I will discuss in \S\ref{sec:nonrelativisticcase}. 
 Also, where W$\&$M use $G_{ab}$ for the force, I use $G$ for geometry and $F$ for universal effects, with $F_{ab}$ as its tensorial form.}  
\setcounter{prop}{1}
\begin{prop} \label{prop:relativisticcase}
    % $(\neg~\forall Curved~\forall Curved)$: 
    [\textbf{The relativistic case}]—Let ($M$, $g_{ab}$) be a relativistic spacetime, let $\tilde{g}_{ab}= \Omega^2 g_{ab}$ be a metric conformally equivalent to $g_{ab}$, and let $\nabla$ and $\widetilde{\nabla}$ be the Levi-Civita
derivative operators compatible with $g_{ab}$ and $\tilde{g}_{ab}$, respectively. Suppose $\Omega$
is non-constant. Then there is no tensor field $F_{ab}$ such that an arbitrary curve $\gamma$ is a geodesic relative to $\nabla$ if and only if its acceleration relative to $\widetilde{\nabla}$ is given by $?F^a_n? \tilde{\xi}^n$, where $\tilde{\xi}^n$ is the tangent field to $\gamma$ with unit length
relative to $\tilde{g}_{ab}$. 
{\normalfont (\cite{WeatherallManchak2014}, p.~242)}
\end{prop}

 \noindent That is, given any geometry, we cannot always write down a new geometry such that a rank-2 tensor relates their geodesics. 
   Proposition 2 is then proven to be correct (in Appendix~\ref{sec:appendixprooftorsion} one can follow the steps of this proof and the use of assumptions, in the context of the logically stronger proposition given in \S\ref{sec:generlisetotorsion}). Thus it is in general not possible to construct empirically-equivalent models by postulating such an $F_{ab}$.

All proofs make assumptions. 
 Evaluating the importance of the W$\&$M proof must involve an investigation of what exactly is assumed,   
 and how weak or strong these assumptions are.
 Dürr $\&$ Ben-Menahem (\cite*{DandbBM2022}) have listed some of the assumptions used in the proof, on which I will largely rely.  
  Yet, I add to this list some topological assumptions and give a finer rendition of the assumption of Riemannian geometry. Also, since W$\&$M explicitly restrict themselves to models of GR, I distinguish between assumptions within GR and assumptions that bring us beyond it.   
 %, a distinction I believe to be crucial for the debate. 
% Regarding the latter, W$\&$M are clear about restricting themselves to GR, and thus for both conceptual reasons as well as reasons of fairness, we should keep apart the leeway for conventionalism one can make within GR and assumptions that bring us beyond the theory.   
% Any criticism of making assumptions that are shared with the theory itself would therefore be unfair and hence I split up the list into assumptions that are shared by GR itself and those that go beyond it. 
  The list is: %\footnote{I assume the internal consistency of GR---call it (\textbf{GR-CONS}).
%Likewise, I assume its empirical adequacy---call it (\textbf{GR-EMP}). The latter may appear equally natural as the former but does involve specifying how exactly the coordination between curves on a manifold and trajectories of physical bodies is operationalised in different epistemic practices. \label{footnote:GR}}:
\begin{itemize}
    \item (\textbf{CONF})  The alternative metric is conformally related to the standard metric $\tilde{g}_{ab}=\Omega^2(x) g_{ab}$, for $d\Omega \neq 0$ (D$\&$BM, p.~156, \textit{non-constancy clause added}). 
    \item (\textbf{NORM}) $\tilde{\xi}^a$ is a unit-norm (with respect to the \textit{new} geometry's metric) tangent vector to the particle's curve: $\tilde{g}_{ab}\tilde{\xi}^a\tilde{\xi}^b=1$ (\textit{ibid.}). 
    \item (\textbf{FORCE}) The [geometrical] alternative to standard acceleration must take the `standard' force-law form $\xi^b\nabla_b\xi^a=\tilde{\xi}^b\widetilde{\nabla}_b\tilde{\xi}^a-?F^a_b?\tilde{\xi}^b$ (\textit{ibid.}, \textit{sign of $?F^a_b?$ adjusted}).
  \end{itemize}
  
\noindent And here are assumptions that would take one beyond GR:  
 % and which struck me as particularly common
% in this field of mathematical physics or uninteresting to change in the current context. Why }:
\begin{itemize}
    \item (\textbf{RIEM}) Geometric alternatives must employ Riemannian geometry: being solely expressible in terms of a metric $\tilde{g}$ and associated Levi-Civita connection $\widetilde{\nabla}$. (\textit{ibid.}) 
    
\noindent    \noindent Here I would add two assumptions constitutive of (\textbf{RIEM}):
    \begin{itemize}
        \item (\textbf{RIEM-SYMM}) the affine-connection is torsion-free.
        \item  (\textbf{RIEM-COMP}) the affine-connection is fully metric compatible, so that the disformation vanishes identically.
    \end{itemize}
     \item \textbf{(TOPO)} Geometric alternatives are constructed on the same manifold. In particular\footnote{As always in philosophy, some assumptions remain suppressed, e.g., smoothness (\textbf{SMOOTH}), paracompactness (\textbf{PARA}) and those that go deeper down.  
     I assume the internal consistency of GR---call it (\textbf{GR-CONS}), and its empirical adequacy---call it (\textbf{GR-EMP}).
 D$\&$BM may very well have regarded (\textbf{TOPO}) as deeper down, so this is meant as supplementation, not correction. For reasons why see \S\ref{sec:programmatic}.}: 
    \begin{itemize}
       \item (\textbf{DIM4}) the manifold in question is restricted to four dimensions. %$a=\curlbrack{0,1,2,3}$.
      \item (\textbf{HAUS}) points can be kept apart by open sets (for every pair of distinct points $p,q \in \mathcal{M}$ there exist neighbourhoods disjoint open sets $U, V \in \mathcal{M}$ such that $p\in U$, $q\in V$ and $U \cap V=\emptyset$). (Hausdorff condition)
    \end{itemize}
\end{itemize}
% Here, (\textbf{RIEM-SYMM}), (\textbf{RIEM-COMP}) and (\textbf{DIM4}) have been added by me.
\noindent These are all required---see W$\&$M's original (\cite*{WeatherallManchak2014}) and Appendix~\ref{sec:appendixprooftorsion} for where exactly.

Generally, (\cite{DandbBM2022}) is a \textit{tour de force} of the history of geometric conventionalism, but D$\&$BM begin their paper by scrutinizing W$\&$M’s proof by questioning its assumptions -- without disputing its formal validity %\footnote{Next to undermining Proposition 2, D$\&$BM pursue two more goals: to articulate their understanding of conventionalism, claiming to align themselves with Poincaré, as well as to reject a distinction between conventionalism and geometric empiricism (often called \textit{physical geometry} (\cite{Helmholtz1866}, cf.~Dewar, Linneman, and Read \cite*[p.~2]{Dewar2022-DEWTEO-11}), which is the view that the geometry of real space is not \textit{a priori} (as Kant held) or hypothetical (as Riemann held) but factual, i.e., empirically accessible by using rigid bodies.}: %This paper touches on touching on all three, this following will focus on the third goal
% }
% , offering rich commentary on holism, underdetermination, pragmatic vs. empirical justification, and the influence of axiomatics and neo-Kantianism in the late 19th and early 20th centuries. 
%  \begin{quote}
% Of course, W$\&$M are aware of the fact that the strength of their theorems on which their reasoning rests depends on their assumptions. But the reader is led to believe that those assumptions are quite natural; to deny them would appear to exact a rebarbatively high price. It is the naturalness of those assumptions (not the validity of W$\&$M's theorems) that we subsequently
% seek to question. (\cite{DandbBM2022}, p.~155)     
%  \end{quote}
 -- by presenting the proof as a \textit{no-go theorem}. 
The above list constitutes a set of mutually inconsistent premises when a particular conventionalist thesis is added: 
% The particular conventionalist premise they add reads:
\begin{itemize}
    \item (\textbf{ALT-ACC}) Geometric alternatives for general-relativistic acceleration of a test-particle, $\xi^b\nabla_b\xi^a$, must exist [...]. (D$\&$BM, p.~155) 
\end{itemize}

\noindent As per usual with inconsistency, 
% one can say that if the proof is correct, 
the proper response is to reject at least one premise, thereby restoring consistency. That is, the conjunction $\neg (\text{(\textbf{ALT-ACC}}) \land \text{(\textbf{CONF})}\land \text{(\textbf{NORM})} \land  \text{(\textbf{RIEM})} \land \text{(\textbf{FORCE})} \land (\textbf{TOPO}))$ 
  is equivalent to the disjunction of the negations of each conjunct: 
$\neg\text{(\textbf{ALT-ACC}})\lor \neg\text{(\textbf{CONF})}\lor \neg\text{(\textbf{NORM})} \lor \neg\text{(\textbf{RIEM})}\lor \neg\text{(\textbf{FORCE})}\lor \neg(\textbf{TOPO})$. 
 In this way, framing a debate in terms of a set of mutually inconsistent premises is a natural way of clarifying and classifying different views in a complex debate, 
by identifying each view with the rejection of one of the premises. (I will articulate such classification in \S\ref{sec:programmatic}).%\footnote{See (\cite{DARDASHTI202147}), who -- as also mentioned in (\cite{Tasdan_Thébault_2023}) -- focuses on the heuristic value of no-go theorems for theory development in physics. Note that this method of identifying the structure of thesis rejection is common in philosophy, both heuristically and for the purposes of classifying by matching positions in an extant debate with the denial of each premise (cf.~(\cite{Haggqvist2009-HGGAMF}; \cite{Mulder2023-MULMRO}) for modal-logical no-go theorems that undermine destructive thought experiments).}) 

% Thus,  
  % This leads D$\&$BM to conclude that 
% \begin{quote}
%     Given these background assumptions, how convincing is W$\&$M's argument? Our answer is deflationary: their reasoning presupposes restrictive, formal premises, unwarranted by conventionalism \textit{as they themselves understand it}. (\cite{DandbBM2022}, p.~155, \textit{original emphasis}) 
% \end{quote}
%  \begin{quote}
% Of course, W$\&$M are aware of the fact that the strength of their theorems on which their reasoning rests depends on their assumptions. But the reader is led to believe that those assumptions are quite natural; to deny them would appear to exact a rebarbatively high price. It is the naturalness of those assumptions (not the validity of W$\&$M's theorems) that we subsequently seek to question. (\cite{DandbBM2022}, p.~155)     
%  \end{quote}

 D$\&$BM attribute to W$\&$M the position $\neg\text{(\textbf{ALT-ACC}})$, that a trade-off between geometries and universal forces is \textit{never} possible, while themselves rejecting one or all of the premises while retaining (\textbf{ALT-ACC}) and their conventionalism.  
  \begin{quote}
Of course, W$\&$M are aware of the fact that the strength of their theorems on which their reasoning rests depends on their assumptions. But the reader is led to believe that those assumptions are quite natural; to deny them would appear to exact a rebarbatively high price. It is the naturalness of those assumptions (not the validity of W$\&$M's theorems) that we subsequently seek to question. (\cite{DandbBM2022}, p.~155)     
 \end{quote}
% Taking such a no-go theorem approach and scrutinising the plausibility of the individual assumptions leads to a rich summary of many extant debates. 
\noindent  They (\cite*{DandbBM2022}, pp.~156-157) continue to interpret (\textbf{NORM}) as ``unwarranted", (\textbf{CONF}) as overly restrictive and ``defeating its purpose", and of ``by-passing" and ``short-circuiting" the proof itself by rejecting such assumptions. 
Denying (\textbf{CONF}) opens up space for underdetermination of models, because -- interpreting conformal rescalings as passive rather than active transformations -- two models may differ by running units, 
where the choice of numerical units trades off against different conformal scalings (\textit{ibid}., p.~157); 
  one can do the same with volume elements (\textit{ibid}., p.~158). 
 By denying (\textbf{NORM}), we can use reparametrisations of curves to create empirically equivalent models that differ with respect to lengths of vectors (\textit{ibid}., p.~156). By denying (\textbf{RIEM}), possibilities open up to
  modify the affine-connection so as to include non-Riemannian spacetimes (\textit{ibid}., pp.~159-160). 
 Finally, by denying (\textbf{FORCE}), interpreting $F$ more loosely as an interaction or ``effect'' rather than a strict ``force'' such as the electromagnetic force represented by the Faraday tensor. 
 I will multiply examples in \S\ref{sec:programmatic}. %The latter would likely have been Reichenbach's strategy (\textit{ibid}., p.~160; but see \cite{MulderCONVSYMM} for a different take).
% Thus,  
%   % This leads D$\&$BM to conclude that 
% % \begin{quote}
% %     Given these background assumptions, how convincing is W$\&$M's argument? Our answer is deflationary: their reasoning presupposes restrictive, formal premises, unwarranted by conventionalism \textit{as they themselves understand it}. (\cite{DandbBM2022}, p.~155, \textit{original emphasis}) 
% % \end{quote}
%  \begin{quote}
% Of course, W$\&$M are aware of the fact that the strength of their theorems on which their reasoning rests depends on their assumptions. But the reader is led to believe that those assumptions are quite natural; to deny them would appear to exact a rebarbatively high price. It is the naturalness of those assumptions (not the validity of W$\&$M's theorems) that we subsequently seek to question. (\cite{DandbBM2022}, p.~155)     
%  \end{quote}

 \section{Universality, not existence: most assumptions are not 
% exploitable as 
loopholes to theorem $\theta$} \label{sec:universalitynotexistence}
% \section{What is and what remains to be proven} \label{sec:naturescopetarget}
 
Surely, W$\&$M's paper is written in an anti-conventionalist tone. 
Reading the title of the paper and some of their commentary, they can easily be misread as implying that their result shows that GR does not allow for conventionalism at all. 
That is, as if trade-offs between metrics and universal forces/effects can never be made, or perhaps only in very exotic cases.
 Furthermore, some assumptions of \S\ref{sec:assumptionslisting} are not explicitly discussed, or only mentioned in passing. 
The one assumption that receives extensive attention is (\textbf{FORCE}). 
Take, for example, (\textbf{CONF}), the restriction to conformally equivalent spacetimes. 
In a footnote, they say:
\begin{quote} 
 Note, though, that requiring conformal equivalence only strengthens our results. If the conventionalist cannot accommodate conformally equivalent metrics, then \textit{a fortiori} one cannot accommodate arbitrary metrics; conversely, if Reichenbach’s proposal fails even in the special case of conformally equivalent metrics, then it fails in the case of (arguably) greatest interest." (W$\&$M \cite*{WeatherallManchak2014}, fn.~13, p.~237)
\end{quote}
A few pages later, the same is said about (\textbf{CONF}) in a footnote to Proposition 2 directly: 
\begin{quote}
Again, this restriction strengthens the result. If the proposal does not work even in 
this special case, it cannot work in general; moreover, the special case is arguably the
most interesting.  (W$\&$M \cite*{WeatherallManchak2014}, fn.~22, p.~242) 
\end{quote} 
This may strike one as quaint, even plainly wrong: aiming to debunk conventionalism, discarding a whole class of spacetimes -- \textit{in casu} the conformally inequivalent spacetimes -- does not strengthen but \textit{weaken} the result. Right? 
If you take a subset of the full set of spacetimes, 
one excludes by fiat a whole range of candidate empirically indistinguishable spacetimes: why is one not allowed to look for empirically equivalent models outside of that subset?   
%  Indeed, D$\&$BM (\cite*[fn 21, p.~161]{DandbBM2022}) say W$\&$M's strategy appears to ``consist in targeting one \textit{special case} of geometric alternatives (viz. those associated with conformally related spacetimes) on the grounds of simplicity." 
% That is, 
% the more logical structure you demand, the less malleable the result.\footnote{For now we set aside the remark on ``simplicity'', which has to do with the tensor-rank restriction on the force concept imposed by (\textbf{FORCE}).}  
 
Both D$\&$BM and W$\&$M do not clearly distinguish between two conceivable positions that Proposition 2 can be taken to dispel, namely 
 whether -- with the help of universal forces -- at least one model is equally capable of making the same predictions as a given model, or whether all of them are. 
 The first is an existence claim, like (\textbf{ALT-ACC}). 
 The second is a universality claim, like theorem $\theta$. 
 The crux of the matter is that there are different kinds of model underdetermination that give rise to different kinds of conventionalism: 
\begin{itemize}
    % \item[-] $\text{\textbf{UDT}-}{\forall g \exists \tilde{g}}$ (\textbf{Existence}): 
        \item[-] \textbf{Existence} ($\text{\textbf{UDT}-}{\forall g \exists \tilde{g}}$): 
    For each metric of an adequate model of some spacetime theory there exists at least one distinct metric of a model of that theory that is equally capable of predicting the same observable consequences, given suitable universal effects. 
    % \item[-] $\text{\textbf{UDT}-}{\forall g \forall \tilde{g}}$ (\textbf{Universality}): 
     \item[-] \textbf{Universality} ($\text{\textbf{UDT}-}{\forall g \forall \tilde{g}}$): 
    For each metric of an adequate model of some spacetime theory any other distinct metric of a model of that theory is equally capable of predicting the same observable consequences, given suitable universal effects. 
\end{itemize}
\noindent  The suitability can be fleshed out broadly or narrowly---W$\&$M flesh it out narrowly as (\textbf{FORCE}), while Reichenbach does so under a broad understanding of universal effects fleshed out under an operationalist semantics à la Helmholtz.  
% under a broad understanding of universal effects fleshed out under Helmholtzian measurability conditions -- that is, an operationalist semantics also alluded by D$\&$BM \cite*{DandbBM2022}, p.~161) -- and by conceptualising of physical geometry rather narrowly by taking a metric tensor to be the unique vehicle for directly representing physical geometrical properties. 
Note that ($\text{\textbf{UDT}-}{\forall g \forall \tilde{g}}$) implies ($\text{\textbf{UDT}-}{\forall g \exists \tilde{g}}$), and also note that, in the context of GR, ($\text{\textbf{UDT}-}{\forall g \exists \tilde{g}}$) takes the form of (\textbf{ALT-ACC}). 
Finally, ($\text{\textbf{UDT}-}{\forall g \exists \tilde{g}}$) is a rendering of theorem $\theta$ in terms of formal model underdetermination (ignoring bona fide conventionalist worries about non-factual choices that enable measurements, cf.~(\cite{FlaviaPadovaniReichi})).

From D$\&$BM's formulation of (\textbf{ALT-ACC}), which states that geometric alternatives must \textit{exist}, it is clear they take W$\&$M's target to be the existence claim,   
% since it states that W$\&$M disprove the claim that geometric alternatives must \textit{exist}. 
 % A closer reading reveals that W$\&$M do not target (\textbf{ALT-ACC})---even when all assumptions are fully granted. 
 % The question whether there in general \textit{exist} particular pairs of Levi-Civita connections whose geodesics can be related by some $F_{ab}$, and hence whether ($\text{\textbf{UDT}-}{\forall g \exists \tilde{g}}$) is correct in GR or Newtonian gravity, is never posed by W$\&$M. 
 not Reichenbach's universality claim. 
 % Indeed, W$\&$M do not consider the universality claim directly.   
Instead, what they disprove is that (roughly) the geodesics of
  any given spacetime $(M,g)$ and those of an alternative but \textit{conformally equivalent} spacetime $(M,\tilde{g})$ can be related by a force field satisfying (\textbf{FORCE}). 
  Hence, we should also consider claims with restricted quantifiers ranging over conformally equivalent metrics (to $g$): 
  \begin{itemize}
    % \item[-] $\text{\textbf{UDT}-}{\forall g \exists_{\text{conf}} \tilde{g}}$ (\textbf{Restricted Existence}): 
  \item[-] \textbf{Restricted Existence} ($\text{\textbf{UDT}-}{\forall g \exists_{\text{conf}} \tilde{g}}$): 
    For each metric of an adequate model of some spacetime theory there exists at least one distinct conformally equivalent metric of a model of that theory that is equally capable of predicting the same observable consequences, given suitable universal effects. 
        % \item[-] $\text{\textbf{UDT}-}{\forall g \forall_{\text{conf}} \tilde{g}}$ (\textbf{Restricted Universality}):  
\item[-] \textbf{Restricted Universality} ($\text{\textbf{UDT}-}{\forall g \forall_{\text{conf}} \tilde{g}}$):  
        For each metric of an adequate model of some spacetime theory any other distinct conformally equivalent metric of a model of that theory is equally capable of predicting the same observable consequences, given suitable universal effects. 
\end{itemize}
 \noindent Again, ($\text{\textbf{UDT}-}{\forall g \forall_{\text{conf}} \tilde{g}}$) implies ($\text{\textbf{UDT}-}{\forall g \exists_{\text{conf}} \tilde{g}}$).
 
 As I read it, W$\&$M slide between two targets, i.e., ($\text{\textbf{UDT}-}{\forall g \exists_{\text{conf}} \tilde{g}}$) and ($\text{\textbf{UDT}-}{\forall g \forall \tilde{g}}$). 
On the one hand, ``If the conventionalist cannot accommodate conformally equivalent metrics, then \textit{a fortiori} one cannot accommodate arbitrary metrics", targets  ($\text{\textbf{UDT}-}{\forall g \forall \tilde{g}}$).   
 On the other hand, ``[...] conversely, if Reichenbach’s proposal fails even in the special case of conformally equivalent metrics, then it fails in the case of (arguably) greatest interest", has in mind ($\text{\textbf{UDT}-}{\forall g \exists \tilde{g}}$), plus a defense of (\textbf{CONF}) so that ($\text{\textbf{UDT}-}{\forall g \exists_{\text{conf}} \tilde{g}}$) implies the unrestricted existence claim. Let us consider both readings. 

 D$\&$BM  (\cite*[p.~156]{DandbBM2022}) say that W$\&$M ``believe this restriction [i.e., (\textbf{CONF})] doesn't diminish the argument's generality, since for Reichenbach \textit{any arbitrary} geometry can be upheld [...]." 
 They attribute to W$\&$M for this belief is their remark about the non-conventionality of causal structure, a point first made by Malament (\cite*{Malament1985-MALAMR}): since Reichenbach famously took causal statements as non-conventional, he would presumably hold that conformal structure is factual because it derives directly from causal facts. Thus, on his own terms,  he should be committed to the conformal part of the metric as non-conventional. 
 On this reading, one would indeed hold that ($\lnot\text{\textbf{UDT}-}{\forall g \exists_{\text{conf}} \tilde{g}}$) implies ($\lnot\text{\textbf{UDT}-}{\forall g \exists \tilde{g}}$), since both classes exhaust the physically viable (or rather: factual) differences.  
Yet, few of us are committed to Reichenbach's causal theory of time---and besides, W$\&$M themselves say they are not primarily concerned with Reichenbach's project. 
One could of course read this as a motivating remark why conformally equivalent spacetimes are \textit{interesting}.\footnote{A worthwhile investigation, \textit{contra} (\cite{Malament1985-MALAMR}), is to see whether the terms `conformal structure', `light-cone structure' and `causal structure', which are synonymous in a modern context, might come apart on different readings of factual and conventional content, and to compare these concepts historically.} 
   The argument why ($\text{\textbf{UDT}-}{\forall g \exists_{\text{conf}} \tilde{g}} \rightarrow \text{\textbf{UDT}-}{\forall g \exists \tilde{g}}$) remains unjustified, or at least implicit. 
 This means that ($\text{\textbf{UDT}-}{\forall g \exists \tilde{g}}$) remains a viable option.\footnote{I thank an anonymous reviewer for pointing me to a recent preprint (\cite{RobertsCONV}) claiming a strengthening of W$\&$M's result without (\textbf{CONF}) and (\textbf{NORM}). Yet the result does not straightforwardly imply Proposition 2: 
   it fails to satisfy (\textbf{FORCE}) as written by D$\&$BM because vectors are not renormalised to the new geometry. But this highlights that (\textbf{FORCE}) and (\textbf{NORM}) are not really independent assumptions. One may instead relax (\textbf{FORCE}) by dropping that $\tilde{\xi}^a$ ought to be unit relative to $\tilde{g}_{ab}$, keeping the tangent unnormalized. (\textbf{FORCE'}) then says that the vector field tangent to $\gamma$ chosen (whatever it is) must satisfy the algebraic force law $\xi^b\widetilde{\nabla}_b\xi^a=?F^{a}_{b}?\xi^b$ and Roberts' Theorem 1 would indeed count as a strict generalisation. In what follows I take the stronger reading, presupposing that $\tilde{\xi}^a$ is unit relative to $\tilde{g}_{ab}$. %as the acceleration of a unit tangent vector. 
  %: a residual underdetermination remains between spacetimes differing in magnitudes involving length. 
  % Despite $\widetilde{\nabla}=\nabla$ and $?F^{a}_{b}?=0$, in the new geometry a standard force is still taken to be proportional to acceleration: $\tilde{\xi}^b\widetilde{\nabla}_b\tilde{\xi}^a=?F^{a}_{b}?\tilde{\xi}^b$ (for a $\nabla$-geodesic).    
  % See also \S\ref{sec:subsec:notCONF}--\S\ref{sec:FORCEabc}.   
    %comparing timelike and spacelike vectors  %Reparametrisation might circumvent this, but then the derivation proceeds differently. 
    %even though then a timelike vector in the old geometry may become spacelike in the new one. 
  \label{fn:RobertsCONF} }

The generalization to torsionful spacetimes in Section 4 is one example of the myriad of possible generalization described in Section 6, and so is meant to lead up to it

% Curiously, D$\&$BM are cognisant of this.

Yet, despite D$\&$BM's objections, denying (\textbf{CONF}) -- or any of the assumptions for that matter, except (\textbf{FORCE}) -- will not save theorem $\theta$. For let us consider ($\text{\textbf{UDT}-}{\forall g \forall \tilde{g}}$) as
 an implicit target of the proof.  
 First, Proposition 2 denies ($\text{\textbf{UDT}-}{\forall g \exists_{\text{conf}} \tilde{g}}$) in the context of GR. 
Then, the restricted universality claim implies the restricted existence claim ($\text{\textbf{UDT}-}{\forall g \forall_{\text{conf}} \tilde{g}} \rightarrow \text{\textbf{UDT}-}{\forall g \exists_{\text{conf}} \tilde{g}}$), so we have (by \textit{modus tollens}) that $\lnot$($\text{\textbf{UDT}-}{\forall g \forall_{\text{conf}} \tilde{g}}$). 
In turn, it is clear that ($\text{\textbf{UDT}-}{\forall g \forall \tilde{g}}$) implies ($\text{\textbf{UDT}-}{\forall g \forall_{\text{conf}} \tilde{g}}$), 
so we have $\lnot$($\text{\textbf{UDT}-}{\forall g \forall \tilde{g}}$). 
In this way we can interpret Proposition 2, roughly, as disproving the statement  ``for any pair of (conformally equivalent) spacetimes $(M,g)$ and $(M,\tilde{g})$, there is a force field satisfying (\textbf{FORCE}) that relates their geodesics'', with the parentheses around ``conformally equivalent'' highlighting the redundancy of this qualification.

Another way to see this is to consider the region of GR spacetimes, assuming (\textbf{TOPO}), that are conformally equivalent (\textbf{CONF}), Riemannian (\textbf{RIEM}), and renormalisable (\textbf{NORM}), as illustrated by the solid green region in Figure~\ref{fig:sosts}. W$\&$M refute that any two spacetimes in this region can be related by a “standard” force tensor field $F_{ab}$ satisfying (\textbf{FORCE}). Suppose we deny (\textbf{CONF}); this expands the space to include Riemannian, renormalisable spacetimes regardless of conformal equivalence. But this merely enlarges the region to that marked ‘GR’ in Figure~\ref{fig:sosts}b, with the original subset (Figure~\ref{fig:sosts}a) still contained within it. The same proof that rules out a force field in the narrower region applies to the wider one.  Strictly analogous reasoning applies to the denial of (\textbf{RIEM}), (\textbf{NORM}) and (if one goes beyond GR) (\textbf{TOPO}): the nature of Proposition 2 as an argument against theorem $\theta$ is that it remains true under generalisation.

\section{Generalising to torsionful connections} \label{sec:generlisetotorsion}
One way of generalising W$\&$M's relativistic result in a controlled way is by explicitly rejecting (\textbf{RIEM}) -- more specifically its constitutive premise (\textbf{RIEM-SYMM}) -- by allowing for torsionful connections, i.e. connections that are not necessarily symmetric:
\begin{prop} \label{prop:relativistictorsioncase}
    % $(\neg~\forall Curved~\forall Curved)$: 
    \textbf{The relativistic torsionful case}—Let ($M$, $g_{ab}$) be a relativistic spacetime, let $\tilde{g}_{ab}= \Omega^2 g_{ab}$ be a metric conformally equivalent to $g_{ab}$, and let $\nabla$ and $\widetilde{\nabla}$ be not-necessarily-symmetric
derivative operators compatible with $g_{ab}$ and $\tilde{g}_{ab}$, respectively. 
Suppose $\Omega$ is non-constant. 
Then there is no tensor field $F_{ab}$ such that an arbitrary curve $\gamma$ is a geodesic relative to $\nabla$ if and only if its acceleration relative to $\widetilde{\nabla}$ is given by $?F^a_n? \tilde{\xi}^n$, where $\tilde{\xi}^n$ is the tangent field to $\gamma$ with unit length
relative to $\tilde{g}_{ab}$. 
\end{prop}
In light of the existence of empirically equivalent torsionful models, i.e., the so-called teleparallel equivalent of GR (cf.~\cite{HayashiShirafuji}), the truth or falsity of this proposition promises to be insightful.   
 In Appendix~\ref{sec:appendixprooftorsion} I show Proposition 3 to be true,  starting from a more general difference tensor as given in (\cite{Jensen2005GeneralRW}). 
 Because the additional torsionful terms do not depend on the metric, they are not affected by a conformal transformation, and so the proof goes through largely analogously to the proof of Proposition 2. 
 This generalization to torsionful spacetimes is one of the myriad possible generalisations in \S\ref{sec:programmatic}.

Again it is important to distinguish targets. If the target is the universality claim ($\text{\textbf{UDT}-}{\forall g \forall \tilde{g}}$), then expanding the scope of a universal quantifier does not affect its falsity:  
if the universality claim fails in the original, narrow case, it trivially fails in the broader one too. 
The key point is that for a given torsionful connection, some others cannot be related by a force field $F_{ab}$, thus violating (\textbf{RIEM-SYMM}) but still refuting theorem~$\theta$ under (\textbf{FORCE}). 
That does not mean this result tells us nothing new: Proposition 3 is a genuine generalisation in the sense that it proves that for any torsionful connection, there exists some (non-trivially conformally equivalent) torsionful connection that cannot be related by a force field $F_{ab}$.  
 
%  Again, it is important to distinguish between two targets.  
% If the target is the universality claim ($\text{\textbf{UDT}-}{\forall g \forall \tilde{g}}$), then the generalisation from Proposition 3 to Proposition 2 is trivial: a proof negating a universality claim is not undermined by expanding the scope of the universal quantifier. In this case,  we expanded the scope from GR, with its symmetric Levi-Civita connection, to (conformally related) spacetimes with not-necessarily symmetric connections, and if the universality claim was false for the original case, then it is false in the generalised case as well. 
% The takeaway is that for a given torsionful connection there exist some connections which cannot be related by a force field $F_{ab}$, violating assumption (\textbf{RIEM-SYMM}) yet still invalidating theorem~$\theta$, at least under (\textbf{FORCE}). 
% That does not mean this result tells us nothing new. 
% In fact, Proposition 3 is a genuine generalisation in the sense that it proves that for any torsionful connection, 
% there exists some (non-trivially conformally equivalent) torsionful connection that cannot be related by a force field $F_{ab}$.  

\begin{figure}%[H]
    \centering
    \begin{subfigure}{0.48\textwidth}
        \centering
        \includegraphics[width=\linewidth]{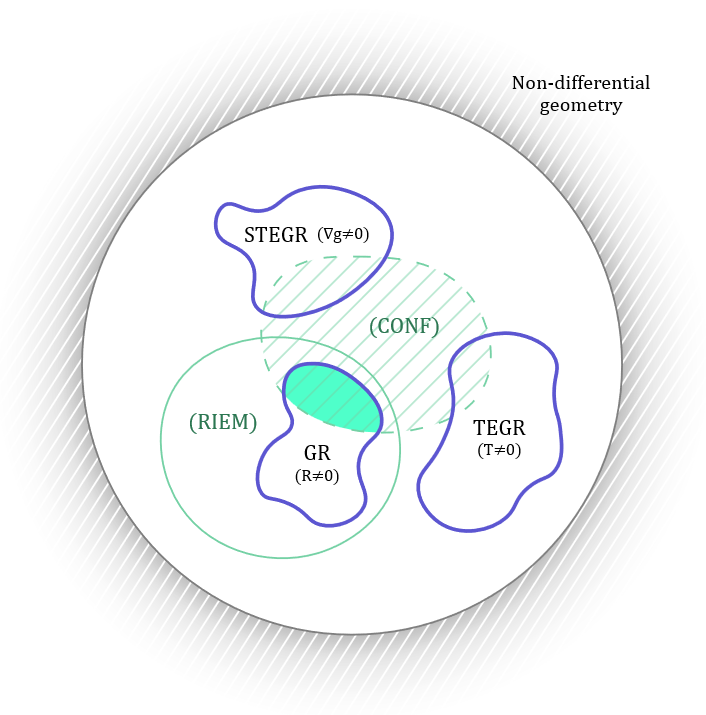}
        \caption{For GR, W$\&$M disprove the existence claim ($\text{\textbf{UDT}-}{\forall g \exists \tilde{g}}$) that there is a force tensor field satisfying (\textbf{FORCE}) that can make two conformally equivalent spacetimes (highlighted region) empirically equivalent. \hfill } 
    \end{subfigure}
    \hfill
    \begin{subfigure}{0.48\textwidth}
        \centering
        \includegraphics[width=\linewidth]{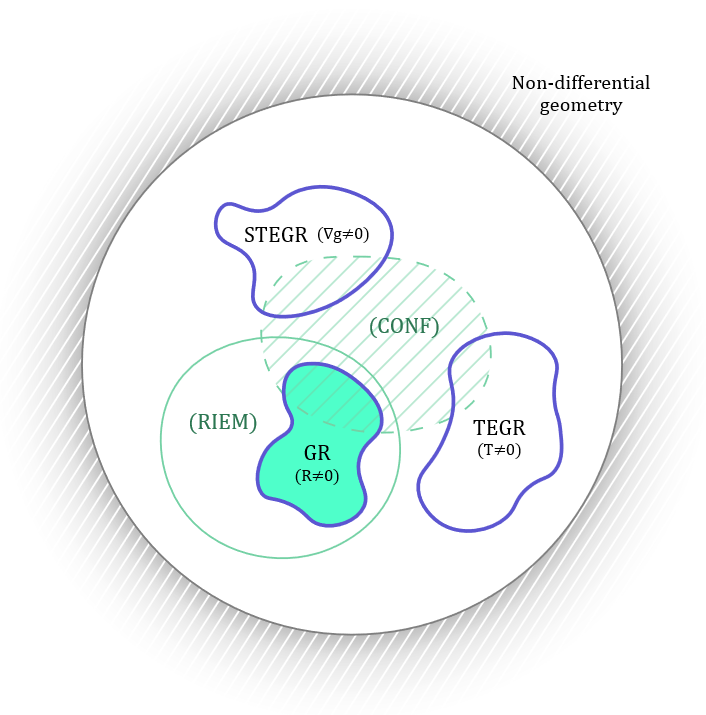}
        \caption{Denying (\textbf{CONF}), one may still believe ($\text{\textbf{UDT}-}{\forall g \exists \tilde{g}}$) could be true in GR. Yet this does not undermine the disproof of the universality claim ($\text{\textbf{UDT}-}{\forall g \forall \tilde{g}}$), but generalises it: theorem $\theta$ remains false under (\textbf{FORCE}).  
        }
    \end{subfigure}
    \vspace{1em}   
    \begin{subfigure}{0.48\textwidth}
        \centering
        \includegraphics[width=\linewidth]{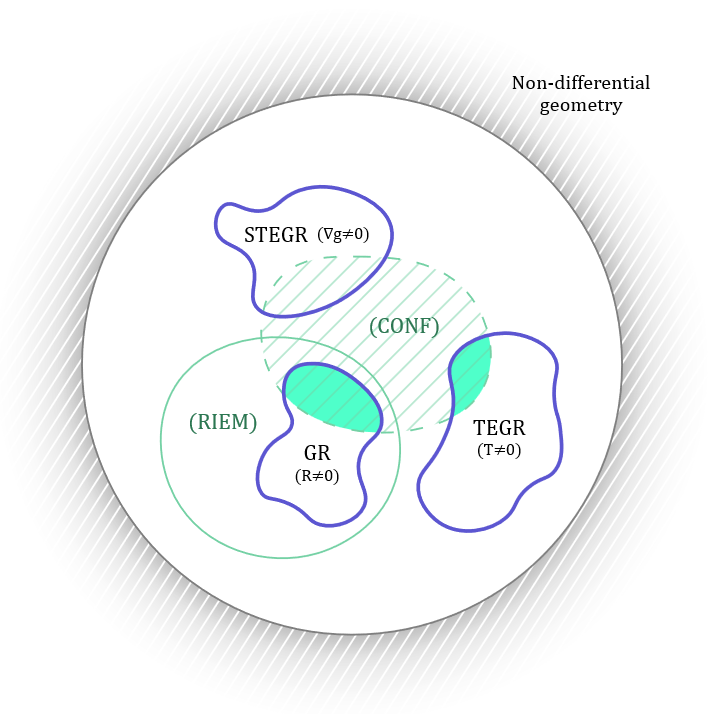}
        \caption{Proposition 3 generalises the result in (a) by allowing torsionful connections (cf.~\S\ref{sec:generlisetotorsion} and Appendix~\ref{sec:appendixprooftorsion}) by rejecting the assumption (\textbf{RIEM-SYMM}). This does not serve as a loophole to save theorem~$\theta$ in GR; more generally, there exists no pair of conformally equivalent torsionful spacetimes related by a tensor field satisfying (\textbf{FORCE}).
        % \hfill \break 
        % \hfill \break
        %  \hfill \break
        % \hfill
        }
    \end{subfigure} 
    \hfill
    \begin{subfigure}{0.48\textwidth}
        \centering
        \includegraphics[width=\linewidth]{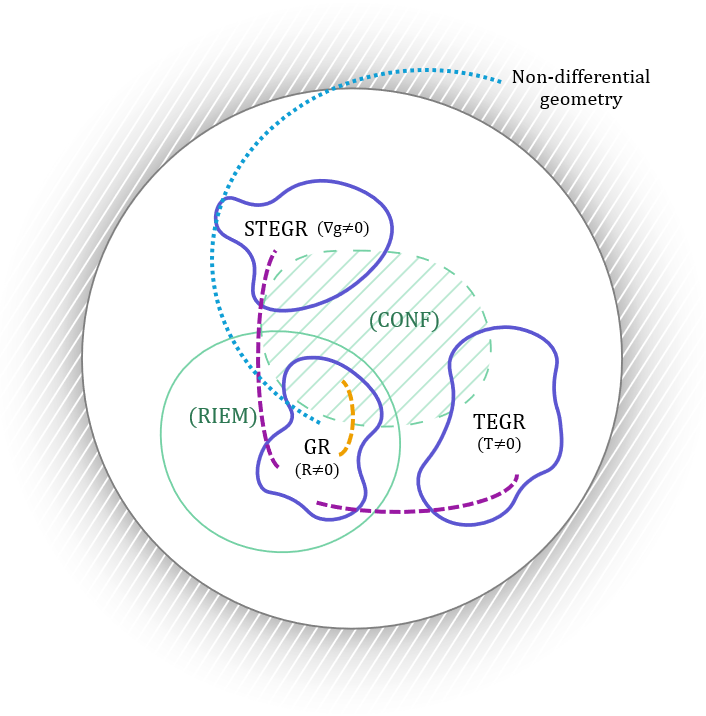}
        \caption{To-be-proven existence claims: the yellow line indicating  ($\text{\textbf{UDT}-}{\forall g \exists \tilde{g}}$), the purple lines the existence claims towards the other two nodes of the geometric trinity of gravity, and the blue line searching for the existence of models equivalent to GR that go beyond differential geometry (e.g. Einstein algebra formulations), all under (\textbf{FORCE}).
 % \hfill \break
  % \hfill 
        }
    \end{subfigure}   
    \caption{\textit{The space of relativistic spacetimes and some inhabitants}.}
    \label{fig:sosts}
\end{figure}

If the target is 
($\text{\textbf{UDT}-}{\forall g~\nabla_{\text{torsionful}}; \exists \tilde{g}~\widetilde{\nabla}_{\text{torsionful}}}$) is taken, 
 this existence claim fails within the class of conformally equivalent spacetimes, see Figure~(\ref{fig:sosts}c). That is, it is a generalisation of Proposition 2 in the sense that for a given torsionful connection there exists no conformally equivalent torsionful connection that can be related by a force field $F_{ab}$.\footnote{I thank an anonymous reviewer for asking me to compare   Proposition 3 to Theorem 1 in (\cite{RobertsCONV}), which considers conformally \textit{in}equivalent spacetimes: one might be able to extend Roberts' Theorem 1 to not-necessarily-symmetric connections. 
 Roberts' conclusion that $?F^a_b? = 0$, however, relies on the symmetry of the connecting field $?C^a_{bc}?$. This fails in the torsionful case. 
 From Theorem 1's proof (\cite*{RobertsCONV}, Eq.~(6)), one sees the torsion must cancel the symmetric part of the connecting field identically: $?C^a_{bc}? = ?K^a_{\brack{bc}}?$. This condition is highly non-generic: a generalisation of Theorem 1 along the lines of Proposition 3 is not straightforward.} %Moreover, as noted in footnote~\ref{fn:RobertsCONF}, Theorem 1 is not strictly a generalisation of Proposition 2, nor is its torsionful counterpart a strict generalisation of Proposition 3.}

To pre-empt misreading: Proposition 3 does not rule out the existence of a teleparallel equivalent to GR. Logically speaking, a force field could still relate torsion-free and torsionful spacetimes, and this should motivate more stringent existence claims (cf.~\S\ref{sec:programmatic}). 
 However, the above framework does allow for a more direct assessment of whether torsion (in teleparallel gravity) counts as a \textit{force}. 
 The torsionful deviation from geodesic motion, governed by the Weitzenböck connection, does not appear to satisfy (\textbf{FORCE}): it enters the geodesic equation via the contorsion tensor contracted with two velocity vectors and cannot generally be cast as a conservative force represented by a rank-2 tensor.   It would be more naturally interpreted as a geometric \textit{effect}, albeit not solely through the metric alone: geometric facts would be represented by the metric \textit{in tandem} with the connection and its associated contorsion. 
In this relativistic context, then, torsion is not a force.

  % Rather, it is better interpreted as a geometric effect—though one encoded not in the metric alone, but in combination with the connection and its contorsion. 

% \footnote{The word `convention', naturally, has many sometimes overlapping but nevertheless distinct meanings that should not be conflated: 
% (a) the colloquial sense as a solution to a multi-actor coordination problem (Lewis \cite*{Lewis1969-Convention}, Ch.~1);
% (b) part of our human judgement that makes -- in the Kantian sense -- something from the world \textit{an sich} comprehensible to us; 
% (c) that which is non-factual, 
% i.e., not synthetic, by
% either (c1) being devoid of truth, 
% or (c2) being analytically true; 
% (d) easily revisable, by being minimally connected to other parts of the theory; 
% (e) choices that fix a gauge. See (\cite{Dorato2009-DOROVS}) for various such senses  and fine-grainings of them. 
%  Here we do not need much fine-graining, since (\cite{WeatherallManchak2014}) only addresses the problem of model underdetermination, not the philosophical position that takes conventions as constitutive principle that make the measurement of physical magnitudes possible in the first place (cf.~\cite{FlaviaPadovaniReichi}). 
% } 

% \FloatBarrier
% \FloatBarrier

 % \FloatBarrier
\section{A programme: systematically exploring the space of spacetime theories} \label{sec:programmatic}

Purely formally, the debate over geometric conventionalism partly stems from the observation that \textit{prima facie} one can trade geometric structure for universal effects, at least mathematically. 
We saw that Reichenbach claimed \textit{any} pair of metrics can be so related and that W$\&$M show that in GR this is not so under (\textbf{FORCE}). Still, ($\text{\textbf{UDT}-}{\forall g \exists \tilde{g}}$) remains neither proven nor refuted, leaving open that \textit{some} metrics might be so related. 
 Were ($\text{\textbf{UDT}-}{\forall g \exists \tilde{g}}$) proven correct for GR, there would definitively be no grounds for the conventionalist view.    
  Unfortunately, it is hard to get mathematical traction on such a general statement as ($\text{\textbf{UDT}-}{\forall g \exists \tilde{g}}$), even under (\textbf{FORCE}). 
  Yet, the formal machinery of Proposition 2 provides intermediate positions, opening up possibilities for articulating -- even generating -- empirically equivalent models, bringing into focus those inhabitants of the formal space of alternative spacetime theories that are known to us.  
   This section explores some options. 

\subsection{Logical structure of the ``go theorem'' and its targets}
Rather than attempting to undermine the proof, I suggest a programmatic attitude: use the list of assumptions in \S\ref{sec:assumptionslisting} not just as inspiration but as formal tools to write down further propositions, quantified either universally or existentially.   
 Tasdan and Thébault (\cite*{Tasdan_Thébault_2023}) adopt a consonant constructive spirit. Citing D$\&$BM, who deny Proposition 2 any merit, Tasdan and Thébault instead give the no-go theorem a more exploratory twist: 
%  \begin{quote}
%   The result of Weatherall and Manchak constitutes a `no-go' theorem for a
% specific form of spacetime conventionalism and, so understood, 
% the theorem might be taken to close off one of the principal avenues for the articulation of such a view within the context of spacetime theory. (\cite[p.~492]{Tasdan_Thébault_2023}) 
% \end{quote}
\begin{quote}
Since the theorem contained in their Proposition 2 is both valid and non-trivial, we take there to be good cause to explore its implications as a ``go theorem'' in the context of the negation of the various physical, mathematical and framework assumptions. (\cite[p.~492]{Tasdan_Thébault_2023})
\end{quote} 
\noindent They then explore five rearticulations of spacetime conventionalism, ``Spacetime Conventionalism 1--5".\footnote{The first three focus on affine structure, inertial structure, and tidal effects, respectively, but are ultimately rejected as either mathematically inconsistent, physically unmotivated, or ruled out by invariant geometric identities. (For further undermining of Conventionalism 3, see Theorem 2 in (\cite{RobertsCONV})). 
Potentially viable theses are Spacetime Conventionalism 4, which concerns the underdetermination of nomic structure, and Spacetime Conventionalism 5, which states that the Bach tensor (i.e., the  conformally invariant part of the Einstein tensor) is not the \textit{unique} conformally invariant rank-2 tensor in four dimensions.}  
Introducing these, Tasdan and Thébault appeal to (\cite{DARDASHTI202147}), which focusses on the heuristic value of no-go theorems for theory development in physics and highlights the fact that no-go results do not dictate \textit{what} must be abandoned, only that \textit{something} in the setup must give.\footnote{Note that this method of identifying the structure of thesis rejection is a common formal method in philosophy, both heuristically and for the purposes of classifying positions of an extant debate by matching positions with denials of premises (see for example (\cite{Haggqvist2009-HGGAMF}; \cite{Mulder2023-MULMRO}) for modal-logical no-go theorems that undermine destructive thought experiments).}   
  I wholeheartedly agree with Tasdan and Thébault's constructive attitude. 
  However, Tasdan and Thébault do not find their five conventionalist theses via the negations of premises.\footnote{That said, 
  for Spacetime Conventionalism 5 Tasdan and Thébault appeal to W$\&$M's no-go theorem to claim that the universal effect tensor $A^{ab}$, i.e., the non-conformally invariant part of the Einstein tensor, is not a standard force apparently because it is not conformally invariant: ``Clearly, by the theorem of Weatherall and Manchak (2014), $A^{ab}$ will not be expressible as a Newtonian force" (\textit{ibid.}, p.~503). The conclusion is correct but the argument is not, because  $lnot$(\textbf{CONF}) is not shown to imply $\lnot$(\textbf{FORCE}). Whether $A^{ab}$ is a force can be evaluated by direct comparison with (\textbf{FORCE}): whether it enters the geodesic equation as a conservative rank-2 force tensor. 
  Because of the contracted Bianchi identity, $\nabla_a G^{ab}=0$, we have $\nabla_a A^{ab}=\nabla_a B^{ab}$, and, because $B^{ab}$ is interpreted to ``characterise facts about geometric spacetime structure" (\textit{ibid.}, p.~503), the divergence $\nabla_a A^{ab}$ is generally not a rank-2 tensor field in the geodesic equation, hence failing (\textbf{FORCE}). 
  Alternatively, if $A^{ab}$ is interpreted as a tidal effect it fails to meet (\textbf{FORCE}) because it does not act on a particle---it represents not the acceleration \textit{of} curves but \textit{between} curves.}
   %  The worry here would be that for many non-vacuum models the alternative geometric construction via $A^{ab}$ would require velocity-dependent forces. Thus one would have to modify the law of energy conservation---running into precisely the problems that Reichenbach's universal effect would. It is not so clear to me how Tasdan and Thébault view the way in which this universal effect can show up in the geodesic equation in a consistent way.  
 % If, alternatively, the universal effect  $A^{ab}$ is taken to act as a tidal effect in the geodesic \textit{deviation} equation, then W$\&$M's assumption of (\textbf{FORCE}) is also violated, for it is then not acting on a particle, modeling the  acceleration of a curve (instead it is modeling the acceleration \textit{between} curves). } 
  Rather, they affirmatively conceive their own alternative conventionalist positions without reference to the logical structure of the theorem. In part, this is because they distinguish -- following (\cite{DARDASHTI202147}) -- between formal possibilities and physical interpretations, to be rejected independently. As such, they characterise (\textbf{RIEM}) as a formal assumption and (\textbf{CONF}) as a physical assumption. 

Below I will proceed differently, staying on a formal level as well as staying closer to the no-go logic of the set of mutually inconsistent premises, actively using the assumptions in \S\ref{sec:assumptionslisting}. 
This is reminiscent of D$\&$BM's approach but gives their systematic approach a new direction along the lines of the constructive spirit of Tasdan and Thébault. It will also cover much additional territory, while keeping an eye on D$\&$BM's requirement that alternatives should at least have a ``modicum of \textit{initial} plausibility'' (\cite*{DandbBM2022}, p.~170), thereby exploring the formal space of spacetime theories that can be generated given GR as a starting point.  
% For this approach opens up possibilities for formulating -- even generating -- empirically equivalent theories, or at least bringing into focus those inhabitants of the formal space of alternative spacetime theories that are known to us.  

\subsection{$\neg$(\textbf{TOPO})}

Assumption (\textbf{TOPO}) is relevant for W$\&$M's proof in that it keeps the topological structure constant: the choice of different connections $\nabla$ and $\widetilde{\nabla}$ is considered to take place on the same manifold, i.e., between $(\mathcal{M},\nabla)$ and $(\mathcal{M},\widetilde{\nabla})$, not between $(\mathcal{M},\nabla)$ and $(\mathcal{N},\widetilde{\nabla})$ (for $\mathcal{N}\neq\mathcal{M}$). 
This is a reasonable assumption, in line with the historical trend of the debate about conventionalism by focussing on local rather than global features. 
However, it is an open question whether some kind of topological conventionality relevantly interacts with traditional geometric conventionality. 

 Consider $\neg$(\textbf{DIM4}). 
In some theories, the effect of what is considered a force in GR can be modelled by a geometric structure in higher dimensions, such as the identification of electromagnetic charge with the value of momentum of particles in the fourth spatial dimension of Kaluza-Klein theory  (\cite{Kaluza}). 
Roberts (\cite*{RobertsCONV}) discusses how simply embedding GR in higher-dimensional spaces from which the four-dimensional spacetime curvature can be recovered generally reflects not a freedom of conventional choice, but creates incompleteness: such frameworks introduce fine-tuning problems and require unexplained features unless completed by additional physical laws. 
Kaluza-Klein theory succeeds \textit{miraculously}, precisely because it posits such laws in five dimensions, eliminating arbitrariness. 

 Next, consider $\neg$(\textbf{HAUS}). 
 That is, the denial that any two distinct points in the manifold can be separated by neighbourhoods. This means that there may be distinct spacetime events that may nevertheless be indistinguishable topologically: points of the manifold become glued together.\footnote{Because showing paracompactness in GR often relies on Hausdorffness, dropping  (\textbf{HAUS}) can make (\textbf{PARA}) harder to establish for some equivalent model of GR. Yet there do exist non-Hausdorff paracompact spaces empirically equivalent to GR (cf.~\cite[Appendix, Lemma $1$]{WuWeatherallForthcoming-WUBASF}).}  
 Luc (\cite*{Luc2020-LUCGMA}) suggests that (\textbf{HAUS}) is not mandatory for GR but usually imposed as a deterministic demand: to rule out future evolutions otherwise  compatible with the same initial data. 
J.B.~Manchak (\cite*[pp.~47--48]{ManchakGlobal}) argues that (non-Hausdorff) Misner spacetime -- where causal regions and CTC regions are neatly separated -- may be physically viable, too. 
 This breakdown of standard GR topology also suggests an algebraic reformulation of spacetime in terms of Einstein algebras (\cite{GerochEinsteinAlgebras}; cf.~\cite{Müller_2013,Rosenstock2015-ROSOEA-2,ShiEinstein}), which avoids the situation that spacetime is defined by points and encodes motion algebraically:  (\textbf{FORCE}) may not even be meaningful.\footnote{I thank an anonymous reviewer for pointing out that, strictly speaking, (\textbf{FORCE}) only risks losing its meaning in non-Hausdorff Einstein-algebra settings where the duality with smooth Lorentzian manifolds breaks down. Within the Hausdorff sector, by contrast, the standard duality guarantees that the notion of ``geodesic" (and hence of a force law) can still be recovered on the algebraic side. Thus, it is the combination of dropping (\textbf{HAUS}) and adopting the Einstein-algebraic framework that threatens the meaning of (\textbf{FORCE}), not the algebraic formulation per se. } %\footnote{Yet, Shi's (\cite*{ShiEinstein}) reconstruction of Einstein algebras in categorical form shows how one can recover manifold‐free ``relational" models via a duality with smooth manifolds. is precisely the kind of algebraic alternative to (HAUS) }

Finally,    
one can also consider upending causal properties by introducing divergent global structures. 
 Reichenbach (\cite*{ReichenbachSpaceTime}, \S12) already discusses underdetermination $G+F+A$ by postulating particular causal anomalies $A$, such as singularities, closed timelike curves, or topological gluing---for an overview of suggestions towards concrete empirically equivalent models $G'+F'+A'$, see for example (\cite[Ch.~5]{ManchakGlobal}) or (\cite{ArntzeniusMaudlin,MulderDieks}) for underdetermination of models in GR with closed timelike curves and models without them.  
 More generally, 
Grimmer (\cite*{Grimmer}) introduces a broad method -- the ISE Method -- to systematically generate topologically distinct but physically equivalent formulations of spacetime theories. The Möbius–Euclid duality developed there illustrates how two theories with very different topologies (e.g., a point particle on a Möbius strip versus a line on the Euclidean plane) can nonetheless encode identical dynamics, thus demonstrating that topological structure can be redescribed without loss of empirical content. ISE potentially provides ample leeway to construct initially plausible topologically distinct alternatives to relativistic models. % Grimmer's paper develops a general framework—spacetime representation theory—for analyzing how distinct topological structures can serve as equivalent representations of the same pre-spacetime theory.

\subsection{$\neg$(\textbf{RIEM})}
It is certainly plausible to deny (\textbf{RIEM}).
W$\&$M pose the underdetermination question within the class of Levi-Civita connections. 
Yet, there is a plausible reason to consider a broader range of affine-connections to inform further propositions, namely  
the existence of metric-affine alternative theories to GR that do not centrally employ Levi-Civita connections. The main examples are the theories of the so-called geometric trinity of gravity. 
In GR, gravitational effects are a manifestation of spacetime curvature, but it is by now well-known there is a theory \textit{prima facie} distinct from GR, 
empirically equivalent to it, and in which gravitational effects are a manifestation of spacetime torsion (cf.~\cite{LyreEynck,Knox2011,MulderRead,WolfReadBoundaries,Weatherall+Meskhidze}). 
This latter theory is known as the teleparallel equivalent of general relativity (TEGR).     Increasingly well-known amongst philosophers (\cite{WolfEquivalence,MulderPSA,WeatherallPSA,ChenRead}) is a third theory (cf.~\cite{HeisenbergReview,Bahamonde}), called the symmetric teleparallel equivalent of general relativity (STEGR), in which gravitational effects are a manifestation of spacetime non-metricity, i.e., the non-compatibility of the connection with the metric.\footnote{Note that these theories generically require a rank-3 ``force" tensor, namely the contorsion   or the distortion tensor, breaking (\textbf{FORCE-b}) (see~\S\ref{sec:FORCEabc}) on top of (\textbf{RIEM}). Thus the geometric trinity is classified in the current taxonomy in the $\lnot$(\textbf{RIEM}) $\land$ $\lnot$(\textbf{FORCE-b}) sector. This is why I concluded in \S\ref{sec:generlisetotorsion} that torsion is not a force in teleparallel gravity. The same will hold for the distortion tensor in non-metric theories.}

Reichenbach had no scruples using an anti-symmetric affine-connection, as becomes clear in his geometrisation of electromagnetism. Giovanelli (Section 4, \cite*{Giovanelli-Appendix}) shows in what way Reichenbach was not naive about these matters, attempting (arguably successfully) to geometrise the electromagnetic field by decomposing the affine-connection into the Christoffel symbols as the product of a mixed anti-symmetrical tensor and a covariant vector. 
% He had a more open mind about the connection, such that the anti-symmetric electromagnetic field can be captured.

Furthermore, the Appendix to \textit{Raum und Zeit} makes clear that Reichenbach saw the geometrisation of gravity as one way of casting the physical content (that is, gravity) in a mathematical mould of geometry, which he regarded as a mere visual ``shiny cloak"---see the recently translated (and long out of print) Appendix to Reichenbach's book analysed in ~(\cite{Giovanelli-Appendix}). 
 This inspires a search for what such a shiny cloak could be hiding, which (less metaphorically) means a ``force theory'' or other causal theory in which the curvature is constant, and in particular flat:
\begin{itemize}
    \item[-] $\text{\textbf{UDT-}}{\forall g \exists \eta}$: for each metric featuring in an empirically adequate model of some spacetime theory, the Minkowski metric  (or Euclidean, for non-relativistic theories)  is capable of reproducing the same observable consequences, given suitable universal effects. %and in particular whether those satisfy (\textbf{FORCE}). 
\end{itemize}
\noindent For GR and under (\textbf{FORCE}), this is proven \textit{false} by Proposition 2, for W$\&$M prove there are models for which this cannot be done, namely for conformally flat models, and hence there is no equivalent flat space standard force version of the theory of GR as a whole.

\subsection{$\neg$(\textbf{FORCE})}\label{sec:FORCEabc}
In fact W$\&$M (\cite*{WeatherallManchak2014}, pp.~235-237) are a bit more detailed than D$\&$BM's formulation of (\textbf{FORCE}). I find three minimal specifications of their standard force: %They specify that a standard force should (\textbf{FORCE}-a) act on a massive body or test particle,  (\textbf{FORCE}-b) be represented by a rank-2 tensor field, and (\textbf{FORCE-c}) be proportional to acceleration (and vanish just in case the acceleration vanishes).
 (\textbf{FORCE-a}) a force is some physical quantity acting on a massive body or point particle;
(\textbf{FORCE-b}) forces are represented by rank-2 tensors at a point; %(at least in Newtonian gravity and general relativity);
 (\textbf{FORCE-c}) the total force acting on a particle at a point must be proportional to the acceleration of the particle at that point, and vanishes just in case the acceleration vanishes.

% However, a universal effect might instead manifest by altering the interpretation of geometric quantities rather than the dynamics themselves. For example, consider a scenario where the actual trajectories remain geodesics of the original connection $\nabla$, but a reinterpretation of which geometric structure represents inertial motion leads one to attribute apparent deviations to an effective universal effect. The causal influence would then lie not in a dynamical term, but in a reinterpretation of which metric or connection governs physical measurements. 
% This is akin to ``absorbing" the effect into a redefinition of the inertial structure, rather than modifying particle dynamics. 

One may deny \textbf{(FORCE-a)} by rejecting that universal effects must enter as force-like terms in the geodesic equation---after all, not all physical influences act by locally deflecting particles from geodesic motion. 
Tasdan and Thébault (\cite*{Tasdan_Thébault_2023}) consider one such route via the geodesic deviation equation, which captures relative acceleration rather than particle-level forces. 
They find it promising but lacking concrete proposals.

One may deny \textbf{(FORCE-b)} by allowing universal effects to be encoded in higher-rank tensor fields. 
%, or in global structures such as holonomies. 
  As a result, one may have to accept that universal effect fail to be conservative. 
 While forces are typically derived from a potential, nothing \textit{a priori} rules out universal effects of a ``drag" type: velocity- or history-dependent interactions familiar from, for example, radiation theory. 
 These violate integrability and energy conservation---serious drawbacks if deemed fundamental, but not obviously ruled out just by empirical adequacy.

To deny \textbf{(FORCE-c)}, one rejects the idea that universal effects must be proportional to acceleration. In other words, that it necessarily manifests as a deviation from inertial motion. In GR, famously, gravity causes no proper acceleration: free-falling particles follow geodesics. Likewise, one might imagine a universal effect that alters relative motion (e.g.\ via curvature or background fields) without pushing particles off their geodesics, and so without acting as a local force in the Newtonian sense.\footnote{
 Finding yourself in a Reichenbachean mood, hypothesise a non-constant scalar field $\psi(x)$ which alters the behaviour of rods and clocks without exerting any force on particles, e.g., shifting atomic frequencies. Test particles would still follow geodesics, but observers using $\psi$-dependent measurement devices observe relative drifts or redshifts. Thus $\psi$ produces observable structure without deflecting motion, evading (\textbf{FORCE-c}).}

 % Instead, the very same trajectories may be reinterpreted as geodesics of a different structure, preserving observable motion while shifting the explanatory burden from dynamics to geometry. In this way, the universal effect alters which metric or connection is physically salient without modifying the equations of motion, i.e. ``absorbing" the effect into a redefined inertial structure rather than modifying particle dynamics.  

\subsection{$\neg$(\textbf{CONF})} \label{sec:subsec:notCONF}
 One may ask whether it is possible to drop the conformal restriction (\textbf{CONF}).  
  % First of all, it is not unthinkable to deny (\textbf{CONF}) on other grounds than those pointed out above (and those pointed out by D$\&$BM). 
 Although today ``conformal structure'', ``light-cone structure'', and ``causal structure'' are used interchangeably, 
one could investigate where such concepts come apart through interpretations that identify their factual or conventional underpinnings differently. 
For Reichenbach, the factual content of the theory of gravity is given by causal relations,  established operationally by means of sending and receiving light signals.  
Before specifying one's method to establish them, lengths and durations are not distinguished.

More to the formal point, under the coordinative definitions W$\&$M adopt, take two arbitrarily related Lorentzian, non-degenerate metrics $g_{ab}$ and $\bar{g}_{ab}$ on $M$, and see whether (\textbf{FORCE}) can be met, assuming also the other assumptions except  (\textbf{CONF})). Is there is a tensor field $F_{ab}$ such that an arbitrary curve $\gamma$ is a geodesic relative to $\nabla$ iff its acceleration relative to $\bar{\nabla}$ is given by $?F^a_n? \bar{\xi}^n$.   
% With the help of the connecting field (Eq.~\eqref{egGeneralDifferenceTensor}) one writes the standard force law in the new geometry as 
% $\tilde{\xi}^b\bar{\nabla}_b \tilde{\xi}^a = ?F^{a}_{b}?\tilde{\xi}^b$, and invoking the connecting field (Eq.~\eqref{egGeneralDifferenceTensor}) on the left-hand side one obtains
 Then introduce the connecting field (Eq.~\eqref{egGeneralDifferenceTensor}) and write the $\bar{g}$–acceleration of a $g$–geodesic, i.e., %by setting $\xi^a=\mathrm{d}x^a/\mathrm{d}\tau$ and $\bar{\xi}^a=Q^{-1}\xi^a$.  One then finds
% \begin{equation} \label{EqGeodesicNonCONF}
  $ \bar{\xi}^b\bar{\nabla}_{b}\bar{\xi}^a
   =\bar{\xi}^b\nabla_b\bar{\xi}^a 
   - ?C^a_{bc}?\bar{\xi}^b\bar{\xi}^c$,  
   %=\cancelto{0}{\xi^b\nabla_b\xi^a} - ?C^a_{bc}?\xi^b\xi^c 
   %- ?C^a_{bc}?\xi^b\xi^c=- ?C^a_{bc}?\xi^b\xi^c.
% \end{equation}
% \begin{equation}
%     \bar{\xi}^b\bar{\nabla}_b\bar{\xi}^a=\bar{\xi}^b\nabla_b\bar{\xi}^a-?C^a_{bc}?\bar{\xi}^b\bar{\xi}^c.
% \end{equation} 
% \noindent 
 where $\bar{\xi}$ is unit with respect to $\bar{g}$. 
 Can the right-hand side reduce to a linear map on the velocity vector field? 
 
 %But %despite the (potentially simplifying) freedom to evaluate this on a $g$-geodesics, 
 Without imposing additional relations, there is no straightforward route to relate $\xi^a$ and $\bar{\xi}^a$. And because covariant differentiation requires knowledge of how a field varies off the curve, there is then no straightforward way to compute $\nabla_b \bar{\xi}^a$. 
  In particular, there is no unique, linear cone-preserving map from $g$-timelike vectors to $\bar{g}$-timelike vectors. 
 This highlights the traction W$\&$M obtained through (\textbf{CONF}): 
if the metrics are conformally equivalent, 
their light-cones coincide, so that there is a one‐to‐one correspondence of the sets of time-like vectors. 
One may attempt to find traction in another way, using a suitable map between the old and new geometries.  
Without that, one is mixing timelike vectors of one geometry with spacelike vectors of the other, turning lengths into durations.

\subsection{$\neg$(\textbf{NORM})}\label{sec:nonorm}
% \textcolor{red}{\textbf{CUT} In the previous subsection, no normalisation was actively imposed. Doing so simply gives $\bar{g}_{ab}\bar{\xi}^a\bar{\xi}^b = g_{ab}(P\bar{\xi})^a(P\bar{\xi})^b = g_{ab} \xi^a\xi^b =1$.  
% Technically: imposing normalisation does not shrink the space of allowable maps but simply picks out those vectors on which you evaluate $?P^a_b?$.} 
 % However, one may believe that dropping (\textbf{NORM}) could help one circumvent the fact that there is no rank-2 tensor field $F_{ab}$ in the non-conformal case.  

  Denying (\textbf{NORM}), i.e. both $g_{ab}\xi^a\xi^b\neq 1$ and $\tilde{g}_{ab} \tilde{\xi}^a\tilde{\xi}^b\neq 1$, the geodesic equation obtains an additional reparametrisation term (cf.~\cite{CarrollBook}, p.~109; \cite{Malament2012-MALTIT}, pp.~58--59):  
\begin{equation}
 \tilde{\xi}^b\widetilde{\nabla}_{b}\tilde{\xi}^a=-?C^a_{bc}?\tilde{\xi}^b\tilde{\xi}^c+\tilde{\xi}^n\widetilde{\nabla}_n\ln({|\tilde{g}_{ab}\tilde{\xi}^a \tilde{\xi}^b|^{1/2})}\tilde{\xi}^a,
\end{equation}
  because $\tilde{\xi}^a$ is no longer unit relative to $\tilde{g}$, so one should correct for the failure of affinity. 
To see this, note that when you reparameterize a geodesic $\gamma(\lambda)$ with tangent $\tilde{\xi}^a = dx^a/d\lambda$ with a new parameter $\bar{\lambda}$, that new parameter generally does not increase at a constant rate relative to the original affine parameter $\lambda$ (unless $d^2\bar{\lambda}/d\lambda^2=0$). As a result, the ``squared speed'' of the curve with respect to $\tilde{g}$, namely $\tilde{g}_{ab}\tilde{\xi}^a \tilde{\xi}^b$, no longer remains constant along the curve itself, i.e. its derivative with respect to $\bar{\lambda}$ (or, equivalently, its covariant derivative along $\tilde{\xi}$) is nonzero. 
 Using metric compatibility and the symmetry of $\tilde{g}_{ab}$, we have 
 \begin{equation}\label{EqChangeSquaredSpeed}
 \tilde{\xi}^b \tilde{\nabla}_b (\tilde{g}_{ac}\tilde{\xi}^a\tilde{\xi}^c) = 2\tilde{g}_{ac} (\tilde{\xi}^b \tilde{\nabla}_b \tilde{\xi}^a) \tilde{\xi}^a.  
 \end{equation}

Also, since the reparametrised curve must trace the same path as the original affine geodesic, the new acceleration $\tilde{\xi}^b\tilde{\nabla}_{b}\tilde{\xi}^a$ can only point along $\tilde{\xi}^a$ itself, so we write the non-affine geodesic equation
\begin{equation} 
\tilde{\xi}^b\tilde{\nabla}_b\tilde{\xi}^a = \kappa \tilde{\xi}^a, 
\end{equation} 
for some scalar function $\kappa(\bar{\lambda})$. 
 Comparing this with Eq.~\eqref{EqChangeSquaredSpeed}, we have 
 \begin{equation}
 \tilde{\xi}^b \tilde{\nabla}_b(\tilde{g}_{ac}\tilde{\xi}^a \tilde{\xi}^c) = 2\kappa \tilde{g}_{ac}\tilde{\xi}^a \tilde{\xi}^c, 
 \end{equation} 
  so that 
  \begin{equation}
  \kappa = \frac{1}{2} \frac{\tilde{\xi}^b \tilde{\nabla}_b(\tilde{g}_{ac}\tilde{\xi}^a \tilde{\xi}^c) }{
  \tilde{g}_{de}\tilde{\xi}^d \tilde{\xi}^e }
      = \tilde{\xi}^b \tilde{\nabla}_b \ln|\tilde{g}_{ac}\tilde{\xi}^a \tilde{\xi}^c|^{1/2},
\end{equation}
where the logarithmic term makes explicit the measure of the failure of unit normalization or affine parametrisation in the new geometry. The parameter $\lambda$ ``runs too fast or too slow'' compared to an affine parameter. 
       This fact is used in D$\&$BM's particular case (\cite*{DandbBM2022}, p.~156), while the above is the general case. 
 
Within GR, (\textbf{NORM}) goes hand in hand with (\textbf{FORCE-c}), for a non-affine parametrisation leads to a force term that is not proportional to acceleration---without normalised tangent vectors, one may still satisfy the algebraic form of (\textbf{FORCE}), but in so doing acceleration is no longer a geometric invariant of the curve, becoming parameter dependent. Note that (\textbf{NORM}) also goes hand in hand with (\textbf{CONF}):   when two spacetimes share light-cone structure, one can simply rescale vectors since they are already pointing in the same direction. 
That is, having the conformal relation $\tilde{g}_{ab}=\Omega^2 g_{ab}$ gives a simple conformal rescaling for normalised \textit{time-like} vectors: $\tilde{\xi}^a=\Omega^{-1}{\xi}^a$.    
% Finally, in the homothetic case of a global \textit{constant} conformal factor, if one wants to satisfy (\textbf{FORCE}) one sees that the choice of norm is not independent from the choice of scale: since a Levi–Civita connection determines its metric only up to a constant scale factor, without a norm condition one cannot straightforwardly match accelerations across geometries, and without a fixed conformal class one cannot easily compare norms.   

\section{Discussion: Conjecture $\theta$ and conceiving of alternatives rigorously}

% \section{Underdetermination, conventionalism and Conjecture $\theta$} \label{sec:discussionConvandConj}
Proposition 2 severely restricts the possibilities for model underdetermination in GR. Yet, it does not remove all: D$\&$BM's (\textbf{ALT-ACC}) may still be conjectured to hold, unless each assumption is independently justified. Future work may well show this to be possible.  %---indeed, the responsibility to make things explicit often falls on those who deny underdetermination.  
% Under (\textbf{FORCE}), the proof refutes the universality claim that all metrics can be traded off against each other, while leaving open the existence claim that at least one alternative metric $\tilde{g}$ may exist for a given $g$.  
% Proposition 2 restricts the possibilities for model underdetermination in GR.  
%  Whether this is significant is in the eye of the beholder. 
% The fact remains that,  
% Yet, as it stands, it does not undermine all conventionalist hopes of ($\text{\textbf{UDT}-}{\forall g \exists \tilde{g}}$): the possibility remains open that there is at least one alternative metric $\tilde{g}$ to a given $g$.  
 % but leaves open the existence claim that at least one alternative metric $\tilde{g}$ may exist for a given $g$. 
% As such, D$\&$BM's (\textbf{ALT-ACC}) can still be conjectured to hold, unless convincing justifications are given for each assumption listed in \S\ref{sec:assumptionslisting}.  
  Yet, there is no rich no-go theorem to save ($\text{\textbf{UDT}-}{\forall g \forall \tilde{g}}$) in GR, for only two premises are sufficient for a contradiction: theorem $\theta$ and (\textbf{FORCE}). 
Even without a justification for the other assumptions, the only loophole to save theorem $\theta$ is the restriction that the universal effect should be like a standard force field. 
The benefit of this constraint is that it affords mathematical tractability, but  conventionalists may reject it by denying either that universal effects must enter the geodesic equation, or that they must be represented by rank-2 tensors. 
Reichenbach likely would have rejected the latter. 
% Much more could be said here; this paper has instead focused on the other assumptions.
 
Note how all-encompassing theorem $\theta$ is: it states that for any arbitrary given metric, any other arbitrary metric is equally good, at least as far as observable facts go. What counts as a duration in one geometry may be a length in another; what is a continuous path in one may appear as a particle popping in and out of existence in another. 
 That's why theorem $\theta$ is a version of the universality claim ($\text{\textbf{UDT}-}{\forall g \forall \tilde{g}}$). %, under a broad understanding of universal effects fleshed out under Helmholtzian measurability conditions -- that is, an operationalist semantics also alluded by D$\&$BM \cite*{DandbBM2022}, p.~161) -- and by conceptualising of physical geometry rather narrowly by taking a metric tensor to be the unique vehicle for directly representing physical geometrical properties. 
 
 Reichenbach (\cite*{ReichenbachSpaceTime}, \S8, p.~31--33) states theorem $\theta$ without proof. Extrapolating from Poincaré's (\cite*{Poincaré1891}) intertranslatability of the three constant-curvature geometries -- flat Euclidean, positively-curved Riemannian, and negatively-curved Bolyai-Lobachevsky geometries -- he concludes that ``No epistemological objection can be made against the correctness of theorem $\theta$" (\cite*{ReichenbachSpaceTime}, p.~33).
 Would we see clearer when we think of it as \textit{Conjecture $\theta$}, in need of proof, as W$\&$M do?   
  The answer lies in some subtle interplay between the semantic and epistemic components of realism.
  
Starting from a conventionalist position that assumes that for a (physical) geometric fact to be \textit{meaningful}, thereby fleshing out the semantic component of realism, we first need to stipulate a baseline to calibrate our measuring apparatuses (cf.~\cite{FlaviaPadovaniReichi}). Hence Reichenbach's mention of ``universal deformation of measuring instruments". With this stipulation itself not epistemically determinable, surely no further proof is required: Poincaré's one-to-one equivalence results \textit{trivially} back it up. %\footnote{  
 % I read D$\&$BM (\cite*{DandbBM2022}, p.~170) in this way when they call Reichenbach's view \textit{trivial semantic holism}: one can always define one's terms relative to some background and propagate those definitions throughout a formal system.
 % This triviality explains part of the title of D$\&$BM's paper: that Reichenbach was not ``entirely wrong", but nevertheless quite trivial.}  
 This amounts to a selective semantic anti-realism: there are no non-conventional geometric facts independent to begin with.%\footnote{See (\cite{MilenaMatt,FlaviaPadovaniReichi,Marschall2024}), as well as work by D$\&$BM themselves, who are leading scholars of conventionalism (\cite{Ben-Menahem2006-BENC-2,DUERR2021,DuerrRead2024}).
%} 
 % At times, Reichenbach can also be read in this way, too. %, namely that no underdetermination would arise in the first place because there is no factual difference between models that differ only on the metric tensor and universal effects. 
   In such a case, however, the designation ``theorem'' may be seen as an exaggeration. %Theorem $\theta$ would amount to saying that geometric structure can always be semantically reconstrued as non-geometric structure. 

Another kind of ``conventionalist'' (better: empiricist) emphasises the epistemic over the semantic by putting their hopes on the existence claim ($\text{\textbf{UDT}-}{\forall g \exists \tilde{g}}$): anti-realism needing only \textit{one} instance of underdetermination. 
  This is the kind of geometric anti-realism W$\&$M take as their foil, and D$\&$BM are also committed to it when they formulated (\textbf{ALT-ACC}) as an existence claim. 
   Thus, both W$\&$M's and D$\&$BM's projects are completely distinct from Reichenbach's. 
%\footnote{To reiterate, that is not to say it is a necessary requirement: even without underdetermination, conventionalism can be maintained by arguing that choices are being made in the very formulation of the theory that are not themselves imposed by empirical facts.} 

The above explains why W$\&$M -- and this is heavily criticised by D$\&$BM (\cite*[pp.~161-162]{DandbBM2022}) -- cite many conventionalists but do not engage with their key concepts: coordinative definitions, congruences, truth.  Of course, a good deal more is to be said about the nature of conventions and associated conventionalisms, about a century-worth of literature more. %fleshes out various conventional choices and their conventionalisms.\footnote{``Convention" has various distinct meanings: 
% (a) the colloquial sense as a solution to a multi-actor coordination problem (Lewis \cite*{Lewis1969-Convention}, Ch.~1);
% (b) part of our human judgement that makes -- in the Kantian sense -- something from the world \textit{an sich} comprehensible to us; 
% (c) that which is non-factual, 
% i.e., not synthetic, by
% either (c1) being devoid of truth, 
% or (c2) being analytically true; 
% (d) easily revisable, by being minimally connected to other parts of the theory; 
% (e) choices that fix a gauge. (cf.~\cite{Dorato2009-DOROVS} for more.)  }
   However, little subtlety is needed to evaluate Proposition 2. For W$\&$M only address the formal problem of the existence of model underdetermination. This clarifies (but does not justify) the sliding between ($\text{\textbf{UDT}-}{\forall g \forall \tilde{g}}$) and ($\text{\textbf{UDT}-}{\forall g \exists \tilde{g}}$) discussed in \S\ref{sec:universalitynotexistence}.  %without addressing conventionalism \textit{per se}. 
 % Whether geometric statements are  ``true-by-convention'' is not central to their approach, and they adopt the same coordinative definitions commonly used to link the formalism of GR to the world.  
  %, in particular the view that takes conventions as constitutive principle that make the measurement of physical magnitudes possible in the first place (cf.~\cite{FlaviaPadovaniReichi}).   

Ultimately, then, W$\&$M’s project is best viewed as distinct from (anti-)conventionalism proper. 
Rather, it considers the leeway between our leading theory of gravity and our leading concept of force, while keeping the usual coordinative definitions of GR intact. 
This is much in the spirit of David Malament: 
\begin{quote}
  Philosophers of science have written at great length about the geometric
structure of physical space. But they have devoted their attention primarily
to the question of the epistemic status of our attributions of geometric
structure. They have debated whether our attributions are \textit{a priori} truths,
empirical discoveries, or, in a special sense, matters of stipulation or
convention. 
It is the goal of this chapter to explore a quite different issue --- the role played by assumptions of spatial geometry \textit{within physical theory} [...]. (\cite[p.~405]{Malament-MALGAS}, \textit{original emphasis})
\end{quote} 
\noindent Seen in this light, I suggest continuing W$\&$M's work by interpreting the relativistic proof of Prop.~\eqref{prop:relativisticcase} as the starting point of a research programme, for example through extensions such as Prop.~3 (\S\ref{sec:generlisetotorsion}) and systematically exploring the space that I have begun to classify (\S\ref{sec:programmatic}).   
 This is only a handful of examples, but these can be multiplied by casting our net wider and wider over a space of alternatives, including those that move away from differential geometry, approaching the unconceived. 
  Indeed, Stanford's (\cite*{KyleStanford}) so-called New Induction reminds us that our current theory space is unlikely to be exhausted.  
    Quite distinct from one's realist commitments,  
  systematic rejections of the assumptions to W$\&$M's proof provide constraints that make possible a controlled exploration of the formal landscape underlying the titular 
``Theory of Spacetime Theories" of (\cite{Lehmkuhl2016-LEHTAT-5}): with newly formulated propositions one can trace routes between charted regions of the space of spacetimes, forming a growing atlas of alternative relativistic formulations.

% \clearpage
\setstretch{1.0}
\section*{Acknowledgements}
  {\footnotesize
  First and foremost I thank Neil Dewar, for reading and sharpening the text and spending many hours in front of a whiteboard together. 
I wholeheartedly thank Hasok Chang, Miguel Ohnesorge, Will Wolf, and two anonymous reviewers for generous input and proofreading the text; and  
Henrique Gomes, Milena Ivanova, James Read, and Bryan Roberts, as well as audiences in 
Bristol, London, and Irvine CA, for generous and stimulating conversations about rigorous conventionalism; 
and Harvey Brown and Matt Farr, for acting as dissertation examiners and improving the ideas of the paper. 
I am further grateful to Patrick Dürr and Jim Weatherall for helpful correspondence. 
 Last I am grateful for their financial support from Trinity College, Cambridge, in the form of the ``Tarner Scholarship in Philosophy of Science and History of Scientific Ideas";   
 and to the American Institute of Physics' ``Robert H.G.~Helleman Memorial Postdoctoral Fellowship". 
}

  % \clearpage
%Appendix starts here%
% \part*{Appendices} \addcontentsline{toc}{section}{\protect\numberline{}Appendices}
 \numberwithin{equation}{part}
    \appendix
        \numberwithin{equation}{section}

% \clearpage
\section{Appendix: proof of Proposition 3} \label{sec:appendixprooftorsion}
\setstretch{1.0}
% The proof of Proposition 3 proceeds along the same lines as the proof of the original Proposition 2 in (\cite{WeatherallManchak2014}) but with a more general starting point. There are no additional non-trivial steps, which is to be expected given the nature of the proof as a rejection of a universality claim, in line with the discussion in \S\ref{sec:universalitynotexistence}. 
% That does not mean that this result tells us just as much about underdetermination as the W$\&$M result---
Proposition 3 in \S\ref{sec:generlisetotorsion} is a genuine generalisation in the sense that it says that for any torsionful connection, 
there exists some (non-trivially conformally equivalent) torsionful connection which cannot be related by a force field $F_{ab}$. 
 To prove it, we take as a starting point Steuard Jensen's (\cite*{Jensen2005GeneralRW})  Eq.~(3.1.28), which is the generalisation of the difference tensor $2?C^a_{bc}?=g^{an}\brack{\nabla_ng_{bc}-\nabla_bg_{nc}-\nabla_cg_{bn}}$, to torsionful connections 
 \begin{equation} \label{eqCabcChristoffelTorsion}
     ?C^a_{bc}?=\frac{1}{2}\tilde{g}^{an}\brack{\nabla_n\tilde{g}_{bc}-\nabla_b\tilde{g}_{nc}-\nabla_c\tilde{g}_{bn}
     +\Delta T_{bcn}+\Delta T_{cbn}-\Delta T_{nbc}}, 
 \end{equation}  
for the torsion tensor $?T^{a}_{bc}?$, which measures the anti-symmetric part of the associated connection (in coordinate-language it is given by $?T^{\rho}_{\mu\nu}? :=  2?\Gamma^{\rho}_{\blockbrack{\mu\nu}}?$), and $\Delta T_{abd}$ %$=?T^{a}_{bc}?-\tilde{T}\indices{^{a}_{bc}}$ 
the difference between the torsion tensors associated with $\nabla$ and $\widetilde{\nabla}$.\footnote{The difference tensor relates two connections to each other as they act on a smooth tensor \textbf{$\alpha$} of arbitrary rank (\cite[Proposition 1.7.3, p.~51]{Malament2012-MALTIT}) as:
  \begin{align} \label{egGeneralDifferenceTensor}
     \brack{\widetilde{\nabla}_{m}-\nabla_{m}}?\alpha^{a_1...a_r}_{b1...b_s}?=& 
     ~?\alpha^{a_1...a_r}_{n...b_s}??C^n_{mb_1}? 
     + ... + 
     ?\alpha^{a_1...a_r}_{b1...n}??C^n_{mb_s}?\\ \nonumber 
     &- ?\alpha^{n...a_r}_{b1...b_s}??C^{a_1}_{mn}? 
     -...-
      ?\alpha^{a_1...n}_{b1...b_s}??C^{a_r}_{mn}?.
 \end{align} 
 Note that the standard difference tensor (cf.~\cite{Malament2012-MALTIT}, p.~78, Eq. 1.9.6) requires (\textbf{RIEM-COMP}) and (\textbf{RIEM-SYMM}), which are thus assumed for Proposition 2; here (\textbf{RIEM-SYMM}) is broken explicitly.}   There is an overall minus sign and $\tilde{g}$'s compared to Jensen's Eq. (3.1.28) because we are expressing this in the original geometry rather than the new geometry.  
 Let us group the torsion terms together as a \textit{generalised contorsion tensor}:
% \begin{equation}
   $ 2K_{abc}:=\Delta T_{abc}+\Delta T_{bac}-\Delta T_{cab}$, 
% \end{equation}
which reduces to the usual contorsion tensor if only one connection is torsionful and the other not. %(Note that for the contorsion tensor to exist, both connections need to be metric compatible, and thus (\textbf{RIEM-COMP}) is assumed.)  
% In the coordinate representation, the contorsion tensor assumes its well-known form as the difference between a torsionful (metric-compatible) connection with coefficients $\widetilde{\Gamma}$ and the Levi-Civita connection with coefficients $\mathring{\Gamma}$, namely 
% \begin{align}
%     ?K^\rho_{\mu\nu}? = \widetilde{\Gamma}\indices{^\rho_{\mu\nu}} - \mathring{\Gamma}^\rho_{\mu\nu} 
%     = \widetilde{\Gamma}\indices{^\rho_{\mu\nu}} 
%     -   \frac{1}{2}g^{\rho\omega}     \brack{\partial_\omega g_{\mu\nu} -\partial_\mu g_{\omega\nu}-\partial_\nu g_{\mu\omega}}.
% \end{align}

% The difference tensor relates two connections to each other as they act on an  smooth tensor \textbf{$\alpha$} of arbitrary rank (\cite[Proposition 1.7.3, p.~51]{Malament2012-MALTIT}) as:
%   \begin{align} \label{egGeneralDifferenceTensor}
%      \brack{\widetilde{\nabla}_{m}-\nabla_{m}}?\alpha^{a_1...a_r}_{b1...b_s}?=& 
%      ~?\alpha^{a_1...a_r}_{n...b_s}??C^n_{mb_1}? 
%      + ... + 
%      ?\alpha^{a_1...a_r}_{b1...n}??C^n_{mb_s}?\\ \nonumber 
%      &- ?\alpha^{n...a_r}_{b1...b_s}??C^{a_1}_{mn}? 
%      -...-
%       ?\alpha^{a_1...n}_{b1...b_s}??C^{a_r}_{mn}?.
%  \end{align}

  Taking (\textbf{CONF}), the metric $\tilde{g}_{ab}=\Omega^2(x) g_{ab}$ of a new geometry relates to the old one via a non-constant conformal factor $d\Omega \neq 0$. Then compute Eq.~\eqref{eqCabcChristoffelTorsion} using Malament's (\cite*[p.~82]{Malament2012-MALTIT}) Proposition 1.9.5 (for upper indices we have $\tilde{g}^{cd}
  =\Omega^2 \tilde{g}^{ca}\tilde{g}^{db}g_{ab}
  =\Omega^{2} \Omega^{-4} g^{cd}
  =\Omega^{-2} g^{cd}$):  
 \begin{align} \label{eqWandMcalcCabc}
     ?C^a_{bc}?-?K^a_{bc}?&=
     \frac{1}{2\Omega^2}g^{an}\blockbrack{
     \nabla_n\brack{\Omega^2g_{bc}}
     -\nabla_b\brack{\Omega^2g_{nc}}
     -\nabla_c\brack{\Omega^2g_{bn}}
     } \nonumber \\
     &= 
     \frac{1}{2\Omega^2}g^{an}\brack{
     g_{bc}\nabla_n\Omega^2 + \Omega^2 \cancelto{0}{\nabla_ng_{bc}}
     -g_{nc}\nabla_b\Omega^2 - \Omega^2 \cancelto{0}{\nabla_bg_{nc}}
     -g_{bn}\nabla_c\Omega^2 - \Omega^2 \cancelto{0}{\nabla_cg_{bn}}
     } \nonumber \\
     &= \frac{1}{2\Omega^2}\brack{
     g_{bc}g^{an}\nabla_n\Omega^2 
     -?\delta_{c}^{a}?\nabla_b\Omega^2
     -?\delta_{b}^{a}?\nabla_c\Omega^2
     },
 \end{align}
where the vanishing terms vanish due to metric compatibility (\textbf{RIEM-COMP}).

As in W$\&$M's proof, given any smooth timelike curve $\gamma$, 
if $\xi^{a}$ is the tangent field to $\gamma$ with unit length relative to (the old) $g_{ab}$, 
then the tangent field to (the same curve) $\gamma$ with unit length relative to (the new) metric $\tilde{g}_{ab}$ is given by 
\begin{equation} \label{eqRescalingVector}
    \tilde{\xi}^{a}=\Omega^{-1}\xi^{a},
\end{equation}
which holds due to (\textbf{NORM}): if $g_{ab}\xi^{a}\xi^{b}=1$ and $\tilde{g}_{ab}\tilde{\xi}^{a}\tilde{\xi}^{b}=1$ then $\tilde{g}_{ab}\tilde{\xi}^{a}\tilde{\xi}^{b}=g_{ab}\Omega^2\tilde{\xi}^{a}\tilde{\xi}^{b}=1$.

 Then, from Eq.~\eqref{egGeneralDifferenceTensor}, we see that the difference tensor acts on a vector as $\widetilde{\nabla}_m\alpha^{a_1}=\nabla_m\alpha^{a_1}-\alpha^{n}?C^{a_1}_{nm}?$. 
Now take any \textit{geodesic} \textit{curve} of the old connection $\nabla_a$, 
and compute the acceleration relative to (the new) $\widetilde{\nabla}_a$, which is      
\begin{align} 
    \tilde{a}^a:=\tilde{\xi}^{n}\widetilde{\nabla}_n\tilde{\xi}^{a}
    &= \tilde{\xi}^{n}\nabla_n\tilde{\xi}^{a} - ?C^a_{nm}? \tilde{\xi}^{n}\tilde{\xi}^{m}      
    = \frac{\xi^{n}}{\Omega} \nabla_n \frac{\xi^{a}}{\Omega}
    - ?C^a_{nm}?
    \frac{\xi^{n}}{\Omega} \frac{\xi^{m}}{\Omega} \nonumber \\
    &= \frac{\xi^{n}}{\Omega} \nabla_n \frac{\xi^{a}}{\Omega}
    + \brack{
    \frac{1}{2\Omega^2}
    \brack{
     ?\delta_{m}^{a}?\nabla_n\Omega^2
     +?\delta_{n}^{a}?\nabla_m\Omega^2
     -g_{nm}g^{ar}\nabla_r\Omega^2 
     }
     - ?K^a_{nm}?
    } 
    \frac{\xi^{n}}{\Omega} \frac{\xi^{m}}{\Omega}, \nonumber
\end{align}
where Eq.~\eqref{eqRescalingVector} is used to replace $\tilde{\xi}$'s and  Eq.~\eqref{eqWandMcalcCabc} invoked in the last line. 
Performing the derivatives and bringing terms together,
\begin{align} \label{eqProp2longcalcTorsion}
    \tilde{\xi}^{n}\widetilde{\nabla}_n\tilde{\xi}^{a} 
    +  ?K^a_{nm}? \frac{ \xi^{n} \xi^{m}}{\Omega^2}
    &= \frac{\xi^{n}}{\Omega} \nabla_n \frac{\xi^{a}}{\Omega}
    + \frac{1}{2\Omega^2}
    \brack{
     ?\delta_{m}^{a}?\nabla_n\Omega^2
     +?\delta_{n}^{a}?\nabla_m\Omega^2
     -g_{nm}g^{ar}\nabla_r\Omega^2 
     } \frac{\xi^{n}}{\Omega} \frac{\xi^{m}}{\Omega}
      \nonumber \\
    &= \frac{1}{\Omega^2} \cancelto{0}{\xi^{n}\nabla_n\xi^{a}} 
    - \frac{1}{\Omega^3} \xi^{n}\xi^{a}\nabla_n\Omega 
     + \frac{2\Omega~}{2\Omega^4} 
     \brack{
     ?\delta_{m}^{a}?\nabla_n\Omega
     +?\delta_{n}^{a}?\nabla_m\Omega
     -g_{nm}g^{ar}\nabla_r\Omega 
     }
     \xi^{n}\xi^{m}
     \nonumber \\ 
    &= - \frac{1}{\Omega^3} \xi^{n}\xi^{a}\nabla_n\Omega 
     + \frac{1}{\Omega^3} 
     \brack{
      \xi^{n}\xi^{a} \nabla_n\Omega
     + \xi^{a}\xi^{m} \nabla_m\Omega
     -\cancelto{1}{g_{nm} \xi^{n}\xi^{m}} g^{ar}\nabla_r\Omega 
     } 
     \nonumber \\
    &= \frac{1}{\Omega^3} 
    \brack{-\xi^{a}\xi^{n} 
     + 2\xi^{a}\xi^{n} 
     - g^{an} 
     } 
     \nabla_n\Omega
     \nonumber \\
    &= \frac{1}{\Omega^3} 
    \brack{\xi^{a}\xi^{n} 
     - g^{an} 
     } 
     \nabla_n\Omega,
     \end{align}
where the vanishing term in the second line is due to being on a geodesic of the old geometry, (\textbf{NORM}) is invoked in the third line,
and indices are relabelled in the fourth. 
% Using Eq.~\eqref{eqCabcChristoffelTorsion}, we see that the calculation in Eq.~\eqref{eqWandMcalcCabc} proceeds exactly the same way. The extra term depends on torsion but is independent of the metric. Thus it is not affected by the conformal transformation to the (new) metric $\tilde{g}$: 
%  \begin{align} \label{eqWandMcalcCabcTorsion}
%      ?C^a_{bc}?
%      &= \frac{1}{2\Omega^2}\brack{
%      g_{bc}g^{an}\nabla_n\Omega^2 
%      -?\delta_{c}^{a}?\nabla_b\Omega^2
%      -?\delta_{b}^{a}?\nabla_c\Omega^2
%      }
%      +?K^a_{bc}?.
%  \end{align}

Suppose for contradiction that a tensor field $F_{ab}$ satisfying (\textbf{FORCE}) exists. This should then balance the old geometry (in which $\gamma$ is a geodesic) in the new geometry:
\begin{equation} \label{eqWhataFORCEshoulddo}
    ?F^a_m?\tilde{\xi}^m=\tilde{\xi}^{n}\widetilde{\nabla}_n\tilde{\xi}^{a}. 
\end{equation}
The left-hand side can be rewritten as 
\begin{equation}
    ?F^a_m?\tilde{\xi}^m
    =\frac{1}{\Omega}?F^a_m?\xi^m
    =\frac{1}{\Omega}\tilde{g}^{an}?F_{nm}?\xi^m,
\end{equation}
while the right-hand side of Eq.~\eqref{eqWhataFORCEshoulddo} is known, for it is given by Eq.~\eqref{eqProp2longcalcTorsion}:  
% \begin{align} \label{eqProp2LHS=RHS}
%     \tilde{g}^{an}?F_{nm}?\xi^m 
%     &= \frac{1}{\Omega^2} 
%     \brack{\xi^{a}\xi^{n} 
%      - g^{an} 
%      } 
%      \nabla_n\Omega,
% \end{align}
\begin{align} \label{eqProp2LHS=RHSTorsion}
    \tilde{g}^{an}?F_{nm}?\xi^m 
    = \frac{1}{\Omega^2} 
    \brack{\xi^{a}\xi^{n} 
     - g^{an} 
     } 
     \nabla_n\Omega 
     - \frac{1}{\Omega}?K^a_{nm}?\xi^n\xi^m, 
\end{align}
for any smooth unit timelike tangent vector $\xi^a$ at any point $p$ (for $F_{ab}$ is a tensor). 

 Now pick any point $p$ where $\nabla_a\Omega \neq 0$ 
and choose at that point two arbitrary distinct unit timelike vectors $\mu^a$ and $\eta^a$ and their superposed vector $\zeta^a=\alpha \brack{\mu^a + \eta^a}$, 
renormalised by the scalar $\alpha$ to be unit relative to $g_{ab}$.
Then, Eq.~\eqref{eqProp2LHS=RHSTorsion} applies to both $\mu^a$ and $\eta^a$ individually as well as to $\zeta^a$ directly, after which the results are set equal to each other. That is, 
%Thus, %applying Eq.~\eqref{eqProp2LHS=RHS} to both $\mu^a$ and $\eta^a$ individually, we have
\begin{align} \label{eqSeparatelyTorsion}
     \tilde{g}^{an}?F_{nm}?\zeta^m 
    &= \alpha\brack{\tilde{g}^{an}?F_{nm}?\mu^m+\tilde{g}^{an}?F_{nm}?\eta^m} \nonumber \\
    &=  \frac{\alpha}{\Omega^2} 
     \brack{\mu^{a}\mu^{n} 
     - g^{an} 
     } 
     \nabla_n\Omega
     + \frac{\alpha}{\Omega^2} 
     \brack{\eta^{a}\eta^{n} 
     - g^{an} 
     } 
     \nabla_n\Omega 
     - \frac{\alpha}{\Omega} \brack{\mu^{n}\mu^{m} 
     + \eta^{n}\eta^{m} 
     } ?K^a_{nm}? \nonumber \\
    &=  \frac{\alpha}{\Omega^2} 
     \brack{\mu^{a}\mu^{n} 
     + \eta^{a}\eta^{n} 
     - 2g^{an} 
     } 
     \nabla_n\Omega
     - \frac{\alpha}{\Omega} \brack{\mu^{n}\mu^{m} 
     + \eta^{n}\eta^{m} 
     } ?K^a_{nm}?,
\end{align}
and applying Eq.~\eqref{eqProp2LHS=RHSTorsion} to $\zeta^a$ directly and then using its definition, we have
\begin{align} \label{eqDirectlyTorsion}
     \tilde{g}^{an}?F_{nm}?\zeta^m 
    =& \frac{1}{\Omega^2} 
     \brack{\zeta^{a}\zeta^{n} 
     - g^{an} 
     } 
     \nabla_n\Omega 
     - \frac{1}{\Omega}
      \zeta^{n}\zeta^{m} 
     ?K^a_{nm}?
     \nonumber \\
    =& \frac{\alpha^2}{\Omega^2} 
     \brack{\mu^{a}\mu^{n} 
     + \mu^{a}\eta^{n} 
     + \eta^{a}\mu^{n} 
     + \eta^{a}\eta^{n} 
     - \alpha^{-2} g^{an} 
     } 
     \nabla_n\Omega 
     \nonumber \\
     &- \frac{\alpha^2}{\Omega}
     \brack{
     \mu^{n}\mu^{m} 
     + \mu^{n}\eta^{m} 
     + \eta^{n}\mu^{m} 
     + \eta^{n}\eta^{m} 
     }?K^a_{nm}?.
\end{align}
 After equating Eqs.~\eqref{eqSeparatelyTorsion} and \eqref{eqDirectlyTorsion}, and rearranging, we finally obtain
\begin{align} \label{eqContradictionTorsion}
     \brack{\frac{2\alpha - 1}{ \alpha }}
     g^{an} \nabla_n\Omega
     &=  \blockbrack{\brack{\alpha-1}
     \brack{\mu^{a}\mu^{n} + \eta^{a}\eta^{n} }
     + 2\alpha \mu^{(a}\eta^{n)}  
     % + 2\alpha \eta^{a}\mu^{n} 
     } 
     \nabla_n\Omega \nonumber \\
     &  - \Omega
     \blockbrack{
     \brack{\alpha-1}
     \brack{\mu^{n}\mu^{m} 
     + \eta^{n}\eta^{m}} 
     + 2\alpha \mu^{(n}\eta^{m)} 
    % + 2\alpha \eta^{n}\mu^{m} 
     }?K^a_{nm}?.
\end{align}
\noindent In the case that the torsional terms on the right-hand side somehow cancel the conformal terms, then resulting equation would state the homothetic case that $\Omega$ is constant, which contradicts the assumption that it is non-trivial. If they do not cancel, then 
 Eq.~\eqref{eqContradictionTorsion} \textit{cannot} simultaneously hold for all choices of $\mu^a$ and $\eta^a$:
the left-hand side is always proportional to $\nabla_a\Omega$, and so has a fixed orientation, independent of $\mu^a$ or $\eta^a$. By contrast, the right-hand side contains explicit bilinear combinations of $\mu^a$ and $\eta^a$, including the $K$-dependent terms, whose orientations vary freely as $\mu^a$ and $\eta^a$ vary; thus no ``miraculous'' cancellations can occur that would collapse the right-hand side into the fixed direction of $\nabla^a\Omega$. Contradiction.
 Thus, also in the torsionful case, there is no tensor field $F_{ab}$ that can everywhere relate the geodesics of $\nabla$ and $\widetilde{\nabla}$: Proposition 3 is true.

\renewcommand{\baselinestretch}{1.0}  
 % \clearpage
 % {\footnotesize 
% }
% \bibliographystyle{apalike}
% \addcontentsline{toc}{section}{Bibliography} 
%  \addcontentsline{toc}{section}{\protect\numberline{}Bibliography}
% \renewcommand\refname{Bibliography}
\begin{refcontext}[sorting=nyt]
% \DeclareNameAlias{sortname}{first-last}
% Define a new name format for the first author
\printbibliography%[]
\end{refcontext}

\end{document}